\documentclass[12pt]{iopart}

\usepackage{iopams}
\usepackage{graphicx}
\usepackage{amssymb}
\usepackage{textcomp}
\usepackage[british]{babel}
\usepackage{color}
\usepackage[T1]{fontenc}

\begin{document}

\title[Single-trajectory spectra of scaled Brownian motion]{Single-trajectory
spectral analysis of scaled Brownian motion}

\author{Vittoria Sposini$^{\dagger,\ddagger}$, Ralf
Metzler$^{\dagger}$ and Gleb Oshanin$^{\flat}$}
\address{$\dagger$ Institute for Physics \& Astronomy, University of Potsdam,
14476 Potsdam-Golm, Germany}
\address{$\ddagger$ Basque Centre for Applied Mathematics, 48009 Bilbao, Spain}
\address{$\flat$ Sorbonne Universit\'{e}, CNRS, Laboratoire de Physique Th\'{e}orique
de la Mati\`{e}re Condens\'{e}e (UMR 7600), 4 Place Jussieu, 75252 Paris Cedex 05,
France}
\ead{rmetzler@uni-potsdam.de (corresponding author)}

\begin{abstract}
A standard approach to study time-dependent stochastic processes is the power
spectral density (PSD), an ensemble-averaged property defined as the Fourier
transform of the autocorrelation function of the process in the asymptotic
limit of long observation times, $T\to\infty$. In many experimental
situations one is able to garner only
relatively few stochastic time series of finite $T$, such that practically
neither an ensemble average nor the asymptotic limit $T\to\infty$ can be
achieved. To accommodate for a meaningful analysis of such finite-length data
we here develop the framework of single-trajectory spectral analysis for one
of the standard models of anomalous diffusion, scaled Brownian motion. We
demonstrate that the frequency dependence of the single-trajectory PSD is
exactly the same as for standard Brownian motion, which may lead one to the
erroneous conclusion that the observed motion is normal-diffusive. However,
a distinctive feature is shown to be provided by the explicit dependence on
the measurement time $T$, and this ageing phenomenon can be used to deduce
the anomalous diffusion exponent. We also compare our results to the
single-trajectory PSD behaviour of another standard anomalous diffusion
process, fractional Brownian motion, and work out the commonalities and
differences. Our results represent an important
step in establishing single-trajectory PSDs as an alternative (or complement)
to analyses based on the time-averaged mean squared displacement.
\end{abstract}

\today

\section{Introduction}

The spectral analysis of measured position time series ("trajectories") $X(t)$
of a stochastic process provides important insight into its short and long time
behaviour, and also unveils its temporal correlations \cite{norton}. In standard
textbook settings, spectral analyses are carried out by determining the so-called
power spectral density (PSD) $\mu(f)$ of the process. The PSD is classically
calculated by first performing a Fourier transform  of an individual trajectory
$X(t)$ over the finite observation time $T$,
\begin{equation}
S(f,T)=\frac{1}{T}\left|\int_0^Te^{ift}X(t)dt\right|^2,
\label{PSD1}
\end{equation}
where $f$ denotes the frequency. The quantity $S(f,T)$ for finite observation
times $T$ is, of course, a random variable. The standard PSD yields from $S(f,T)$
by averaging it over a statistical \emph{ensemble\/} of all possible trajectories.
After taking the asymptotic limit $T\to\infty$, one obtains the standard PSD
\begin{equation}
\label{main}
\mu(f)=\lim_{T\to\infty}\frac{1}{T}\left\langle\left|\int_0^Te^{ift}X(t)dt\right|^2
\right\rangle,
\end{equation}
where the angular brackets denote the statistical averaging. 

Following this definition, the standard PSD was determined for various processes
across many disciplines. This includes, for instance, the variation of the loudness
of musical performances \cite{2}, the temporal evolution of climate data \cite{4}
and of the waiting-times between earthquakes \cite{5}, the retention times of
chemical tracers in groundwater \cite{6} and noises in graphene devices \cite{7},
fluorescence intermittency in nano-devices \cite{8}, current fluctuations in
nanoscale electrodes \cite{9}, or ionic currents across nanopores \cite{14}. The
PSD was also calculated analytically for individual time series in a stochastic
model describing blinking quantum dots \cite{18}, for non-stationary processes
taking advantage of a generalised Wiener-Khinchin theorem \cite{20,21}, for the
process  of fractional Brownian motion with random reset \cite{10}, the running
maximum of a Brownian motion \cite{11}, as well as for diffusion in strongly
disordered Sinai-type systems \cite{12}, to name but a few stray examples.

An alternative approach geared towards realistic experimental situations was
recently proposed---based directly on the finite-time, single-trajectory PSD
(\ref{PSD1}) \cite{PSD_BM,PSD_fBM} (see also \cite{pers}). The need for such
an alternative to the standard PSD (\ref{main}) is two-fold. First, while the
asymptotic limit $T\to\infty$ can well be taken in mathematical expressions,
it cannot be realistically achieved experimentally. This especially holds for
typical, modern single particle tracking experiments, in which the observation
time is limited by the microscope's focus or the fluorescence lifetime of the
dye label tagging the moving particle of interest \cite{15}. In general, apart
from the dependence on the frequency $f$ the single-trajectory PSD (\ref{PSD1})
therefore explicitly is a function of the observation time $T$. Moreover,
fluctuations between individual results $S(f,T)$ of the single-trajectory PSD
will be observed, even for normal Brownian motion \cite{PSD_BM}. Second, and
maybe even more importantly, while such fluctuations between trajectories may,
of course, be mitigated by taking an average over a statistical ensemble, in
many cases the number of measured trajectories is too small for a meaningful
statistical averaging. Indeed, for the data garnered in, for instance, \emph{in
vivo\/} experiments \cite{15}, climate evolution \cite{16}, or the evolution of
financial markets \cite{17} one necessarily
deals with a single or just a few realisations
of the process. As we will show, despite the fluctuations between individual
trajectories relevant information can be extracted from the frequency and
observation time-dependence of single-trajectory PSDs. Even more, the very
trajectory-to-trajectory amplitude fluctuations encode relevant information,
that can be used to dissect the physical character of the observed process.

How would we understand an observation time-dependence? This is not an issue,
of course, for stationary random processes, but apart from Brownian motion, only
very few naturally occurring random processes are stationary. A $T$-dependent
evolution of the PSD can in fact be rather peculiar and system dependent. For
instance, the PSD may be ageing and its amplitude may decay with $T$, as it
happens for non-stationary random signals \cite{21}, or conversely, it can
exhibit an unbounded growth with $T$, a behaviour predicted analytically and
observed experimentally for superdiffusive processes of fractional Brownian
motion (FBM) type \cite{PSD_fBM}. As a consequence, the standard textbook
definition (\ref{main}) of the PSD which emphasises the limit $T\to\infty$,
can become rather meaningless.

Motivated by the two arguments in favour of using a single-trajectory approach
to the PSD---the lack of sufficient trajectories in a typical experiment in
order to form an ensemble average and insufficiently long observation times
$T$---references \cite{PSD_BM,PSD_fBM} concentrated on the analysis of the
random variable $S(f,T)$ defined in (\ref{PSD1}) for arbitrary finite $T$ and
$f$. Both for Brownian motion and FBM with arbitrary Hurst index (anomalous
diffusion exponent, see below) a range of interesting, and sometimes quite
unexpected features were unveiled, as detailed in the comparative discussion
at the end of section \ref{sec3}.

While FBM, whose single-trajectory PSD is studied in \cite{PSD_fBM} is a quite
widespread anomalous diffusion process, it is far from the only relevant example
of naturally occurring random processes with anomalous diffusive behaviour. As,
in principle, $S(f,T)$ may behave distinctly for different stochastic processes,
in order to get a general and comprehensive picture of the evolution in the
frequency domain, one needs to study systematically the single-trajectory PSDs
of other experimentally-relevant processes, such as, scaled Brownian motion (SBM),
the continuous time random walk, or diffusing diffusivity models, to name just a
few. In all these examples the microscopic physical processes underlying the
global departure from standard Brownian motion are different, and we would expect
that this difference in the microscopic behaviour translates into the behaviour
in the frequency domain.

Here we concentrate on trajectories $X_{\alpha}(t)$ generated by SBM,
a class of non-stationary anomalous
diffusion processes encoding the mean squared displacement (MSD) $\langle X_{\alpha}^2
(t)\rangle\simeq t^{\alpha}$ with anomalous diffusion exponent $\alpha$. SBM was
formally studied within different contexts in the last two decades \cite{SBM_Lim,
SBM_Sokolov,SBM_Jeon}. Historically, it was introduced
already by Batchelor in 1952 in the context of the turbulent motion of clouds of
marked fluids \cite{SBM_Batchelor}, originally studied by Richardson in 1926
\cite{SBM_Richardson}. An important application of SBM is for particle motion in
the homogeneous cooling state of force-free cooling granular gases, in which the
continuously decaying temperature (defined via the continuously dissipating
kinetic energy) effectively leads to a time-dependence of the self-diffusion
coefficient of the gas \cite{SBM_Metzler}. SBM also describes the dynamics of a
tagged monomer involved into a processes of irreversible polymerisation
\cite{SBM_Moreau}. Similar dynamics emerge in the analysis of fluorescence
recovery after photobleaching (FRAP) data \cite{SBM_Saxton}, as well
as of fluorescence correlation spectroscopy (FCS) data \cite{SBM_Berland}, which
are both widely used techniques to measure diffusion of macromolecules in living
cells and their membranes. Lastly, essentially the same type of anomalous diffusion
modelling was used for the analysis of potential water availability in a region
due to precipitation (snow and rain) \cite{SBM_Porporato}.

The outline of the paper is as follows. In section \ref{sec2} we present the
basics of SBM, introduce our notation, and define the properties under study.
Section \ref{sec3} is devoted to the spectral analysis of single-trajectory
PSDs governed by SBM. Here, we first derive an exact expression for the
moment-generating function of the random variable $S(f,T)$ and evaluate the
exact form of the associated probability density function (PDF). The form of the
latter turns out to be entirely defined by its first two moments, in analogy
to the parental process $X_{\alpha}(t)$. We then present explicit forms of
these two moments, valid for arbitrary anomalous diffusion exponent $\alpha$,
frequency $f$, and observation time $T$. Section \ref{sec3} ends with a
comparative discussion of our results with the behaviour of the single-trajectory
PSD for FBM. Finally, we conclude with a brief summary of our results and a
perspective in section \ref{conc}.

\begin{figure}
\centering
\includegraphics[width=0.49\textwidth]{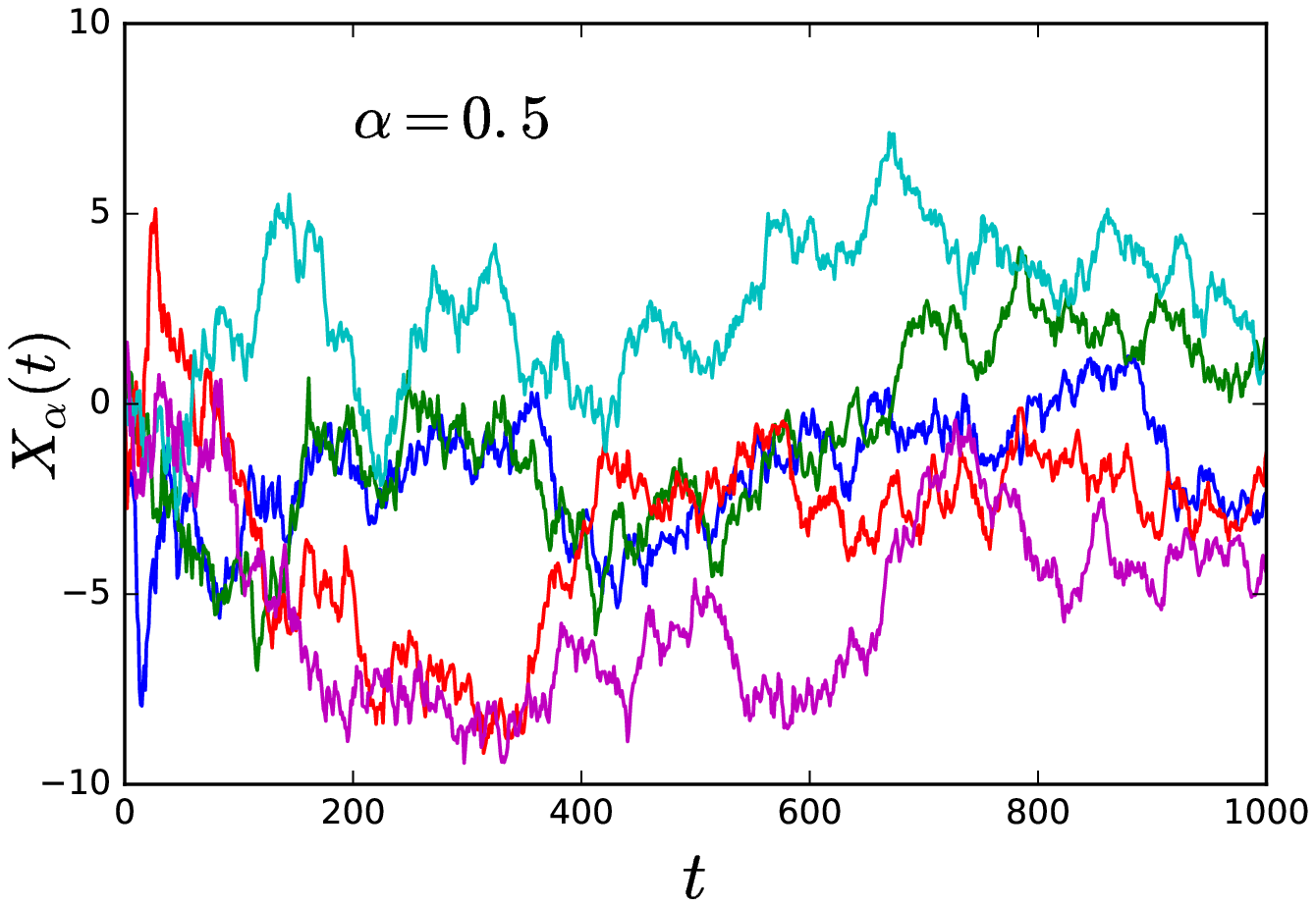}
\includegraphics[width=0.49\textwidth]{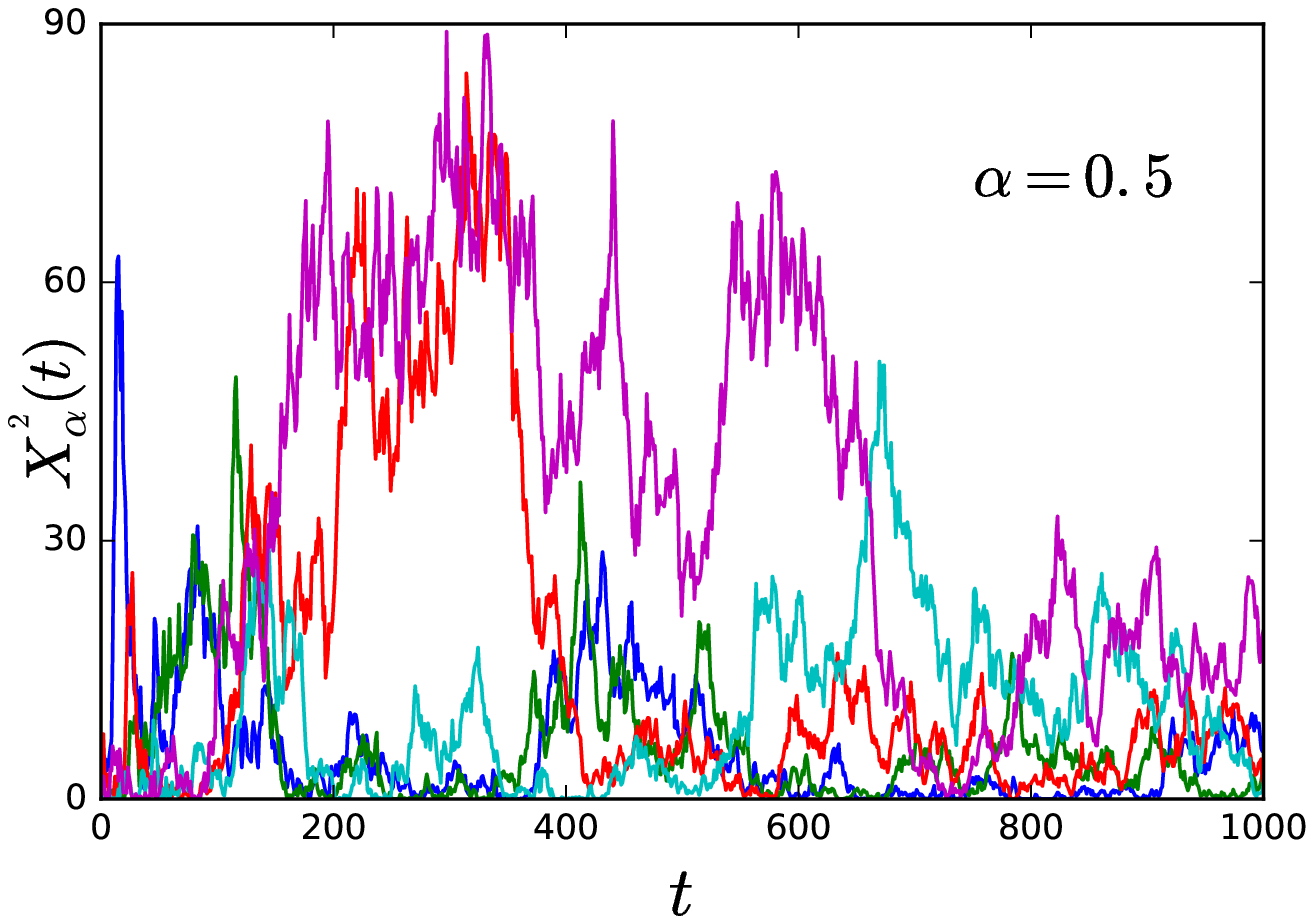}\\
\includegraphics[width=0.49\textwidth]{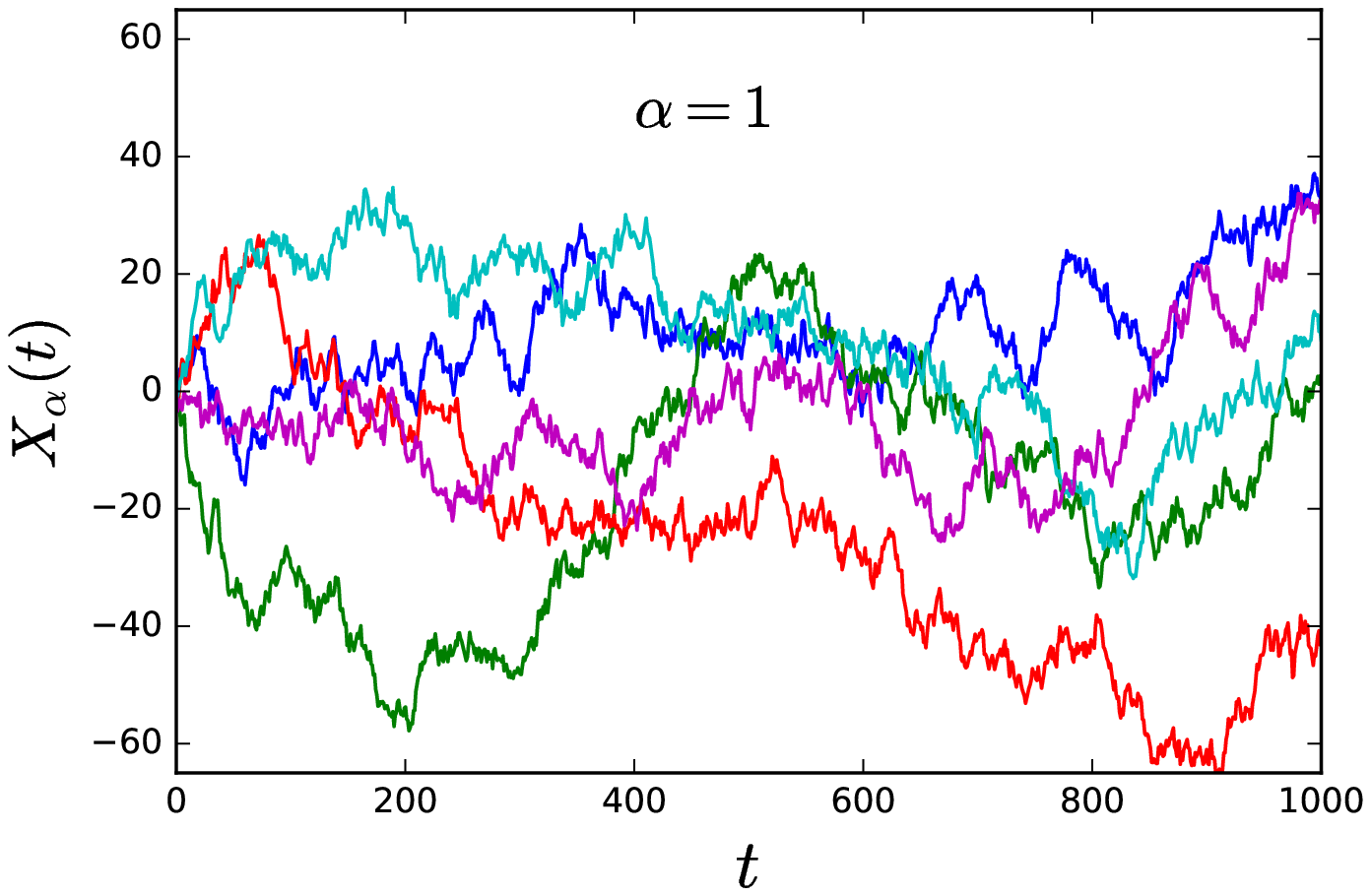}
\includegraphics[width=0.49\textwidth]{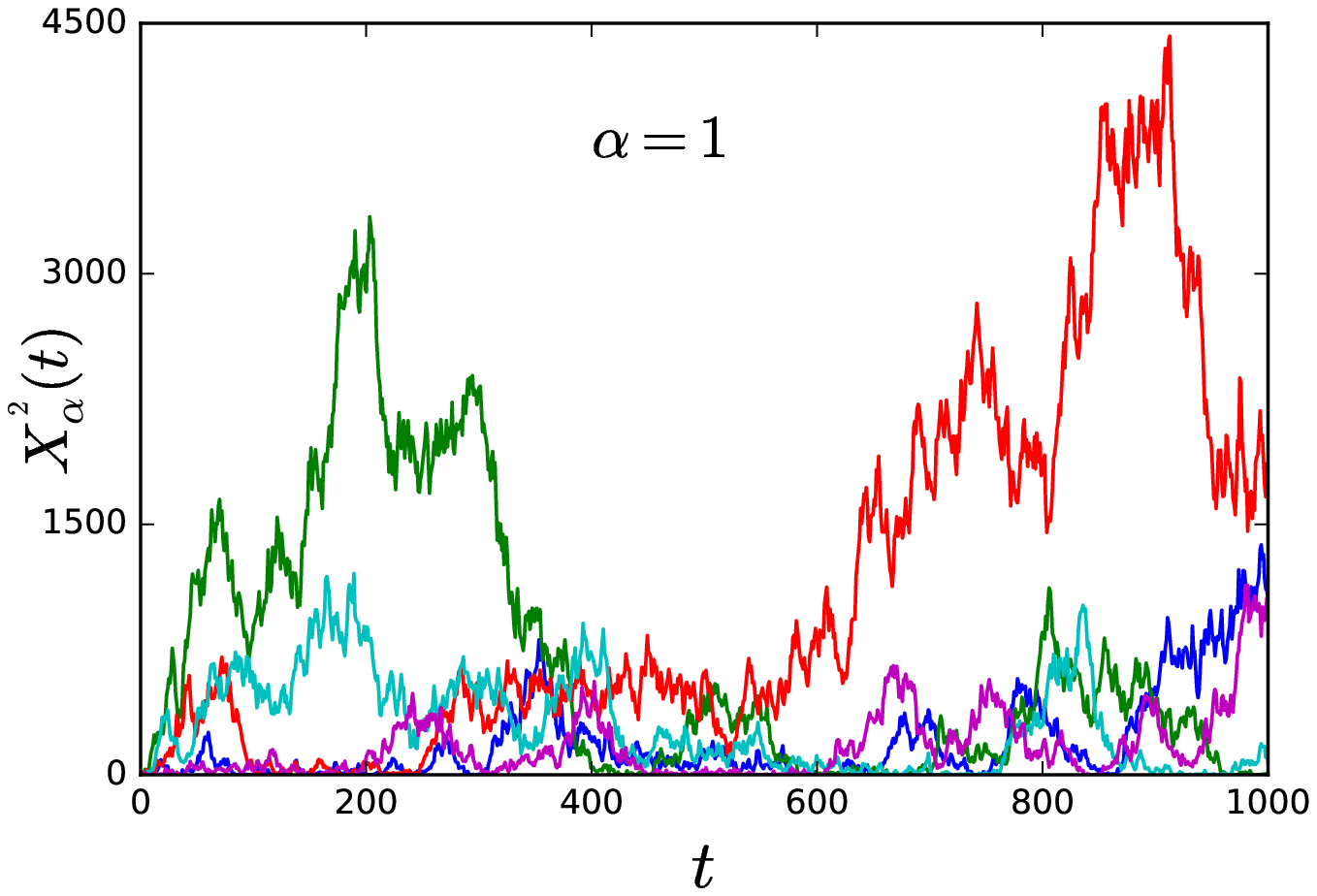}\\
\includegraphics[width=0.49\textwidth]{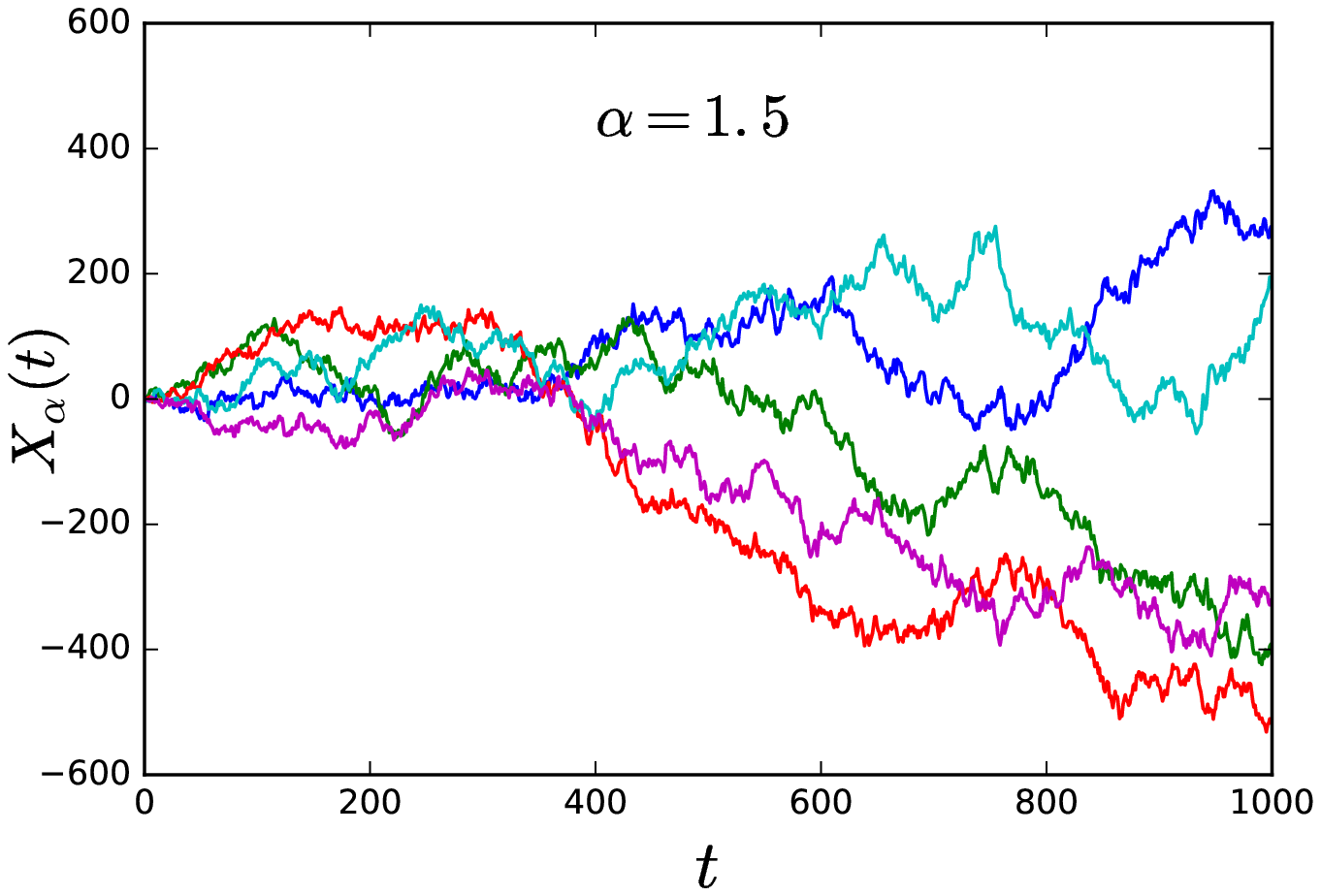}
\includegraphics[width=0.49\textwidth]{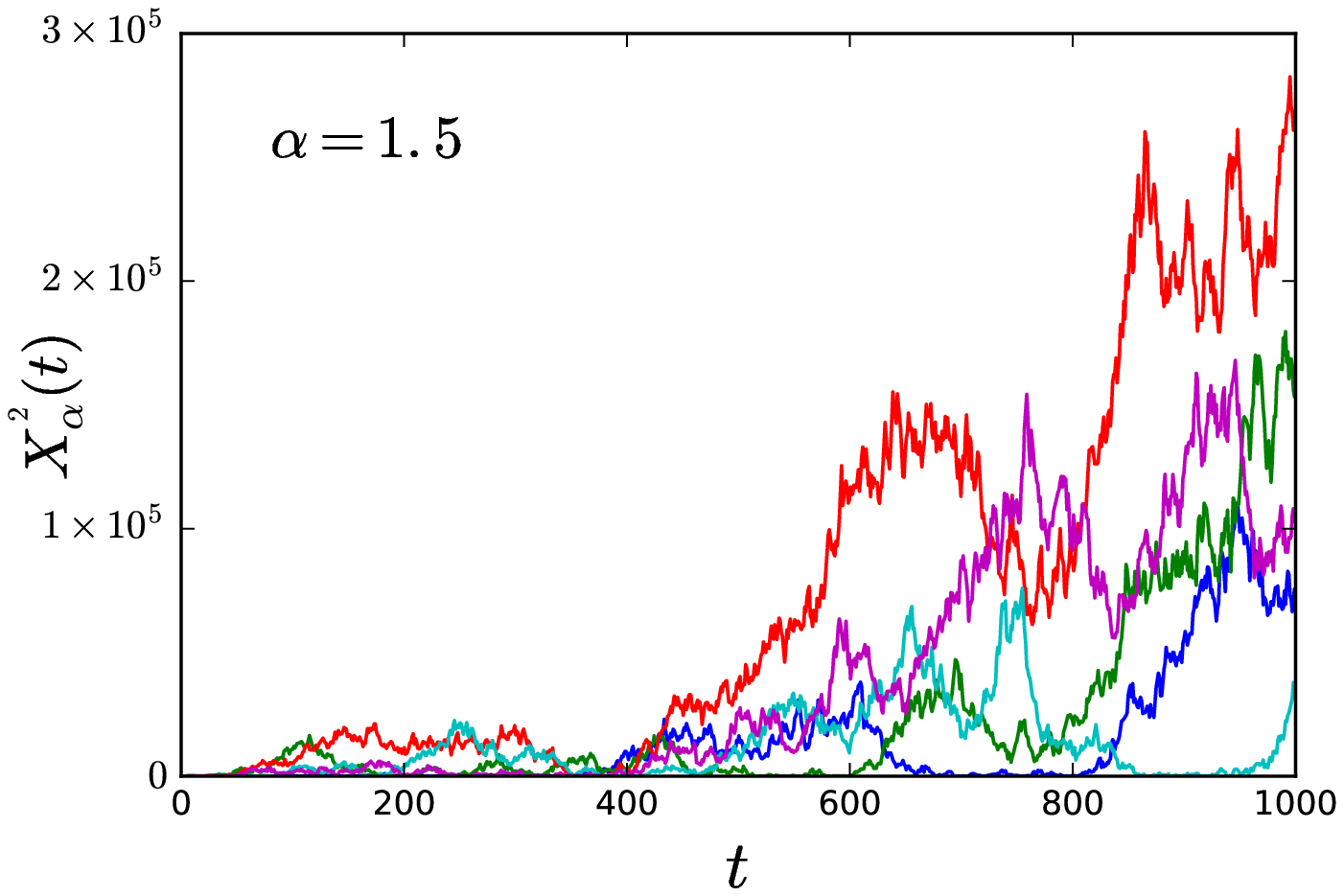}
\caption{Four individual realisations $X_{\alpha}(t)$ for SBM with different
anomalous diffusion exponent for subdiffusion ($\alpha=0.5$, top), normal
diffusion ($\alpha=1$, middle), and superdiffusion ($\alpha=1.5$, bottom).
In the left column we show  the process $X_{\alpha}(t)$ itself, while in
the right column we display its square, $X_{\alpha}^2(t)$.}
\label{img0}
\end{figure}

\section{Model and basic notations}
\label{sec2}

SBM $X_{\alpha}(t)$ is an $\alpha$-parametrised family of Gaussian stochastic
processes defined by the (stochastic) Langevin equation \cite{SBM_Lim,
SBM_Sokolov,SBM_Jeon}
\begin{equation}
\frac{d X_\alpha(t)}{dt}=\sqrt{2D_\alpha(t)}\times\xi(t),
\label{SBM1}
\end{equation}
where $\xi(t)$ denotes Gaussian white-noise with zero mean and the variance
$1/2$, such that 
\begin{equation}
\langle\xi(t_1)\xi(t_2)\rangle=\delta(t_1-t_2),.
\end{equation}
Moreover, $D_\alpha(t)$ is the diffusion coefficient, that follows the
deterministic power-law in time\footnote{The limit $\alpha=0$ corresponds to
the case of ultraslow diffusion with a logarithmic MSD, as studied in \cite{anna}.}
\begin{equation}
D_\alpha(t)=\alpha K_\alpha t^{\alpha-1},\quad 0<\alpha<2,
\label{SBM2}
\end{equation}
where the coefficient $K_{\alpha}$ has physical dimension $\mathrm{cm}^2/\mathrm{
sec}^{\alpha}$. In general, SBM describes anomalous diffusion, such that the
ensemble-averaged MSD scales as a power law in time, 
\begin{equation}
\label{msd}
\langle X_\alpha^2(t)\rangle=2K_\alpha t^\alpha.
\end{equation}
When $0<\alpha<1$ one observes subdiffusive behaviour, while for $1<\alpha<2$
SBM describes superdiffusion. Standard Brownian motion is recovered in the limit
$\alpha=1$. In figure \ref{img0} we depict four representative trajectories of
$X_{\alpha}(t)$ for the subdiffusive, normal-diffusive, and superdiffusive cases.
We note that, especially for the subdiffusive case $\alpha=1/2$ the non-stationary
character is not immediately obvious from the graph of $X_{\alpha}(t)$\footnote{One
may infer the slower spreading rather from comparison of the span of $X_{\alpha}(t)$
on the vertical axis.}, while the character of the process becomes somewhat more
obvious when we plot the square process, $X_{\alpha}^2(t)$. Concurrently, in the
superdiffusive case the growing fluctuations and large excursions away from the
origin appear relatively more pronounced.

\begin{figure}
\centering
\includegraphics[width=0.5\textwidth]{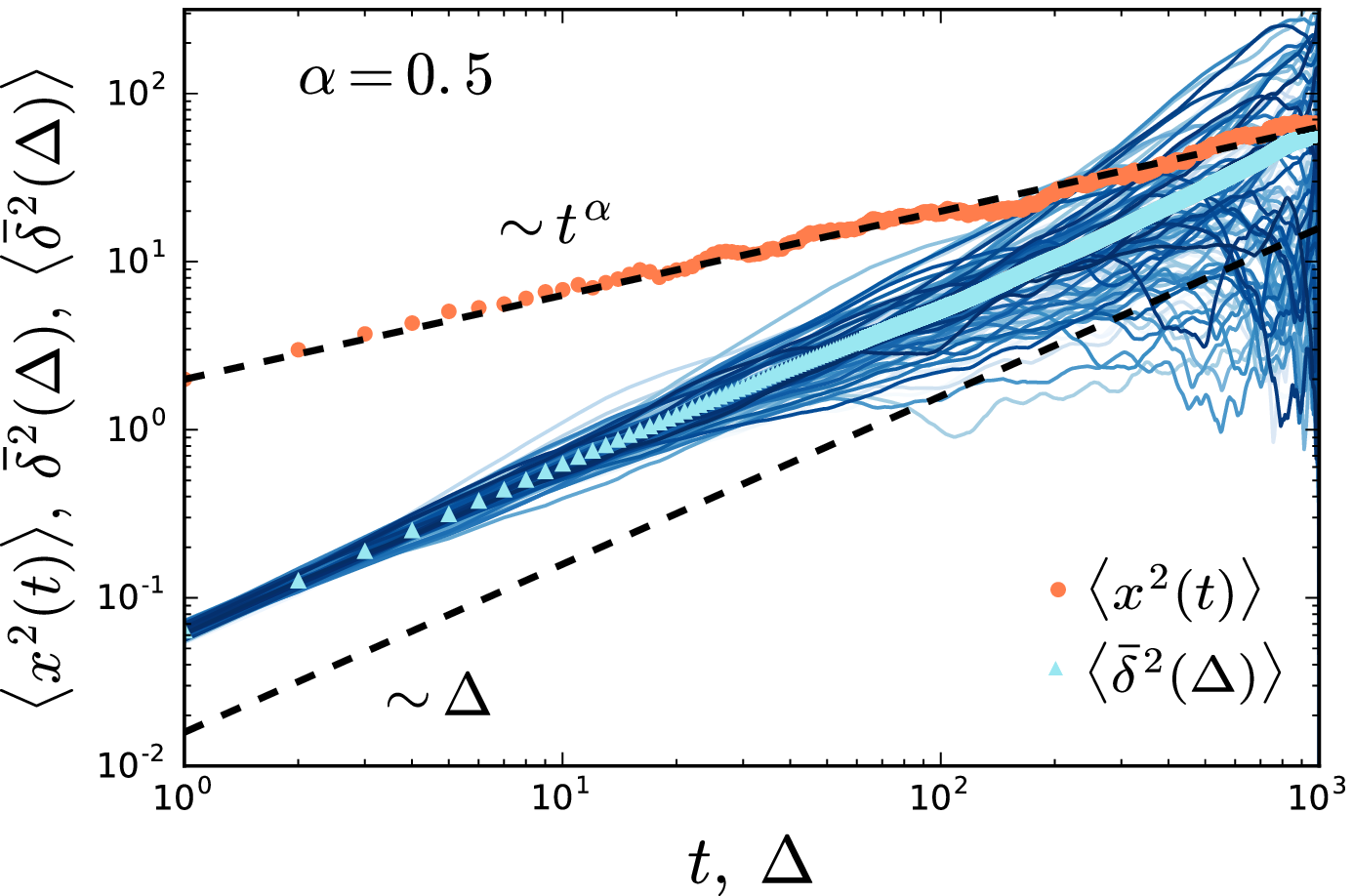}
\includegraphics[width=0.5\textwidth]{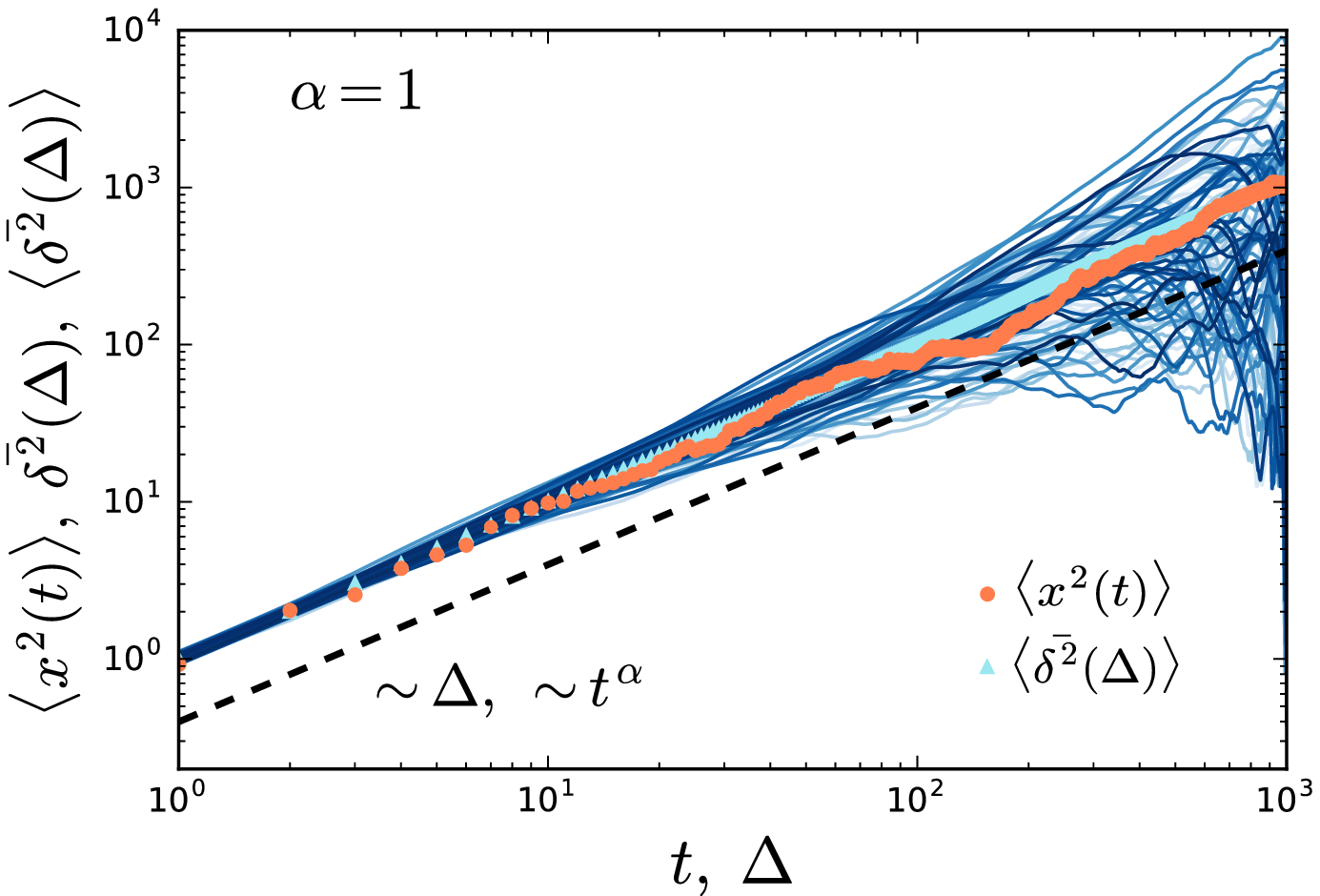}
\includegraphics[width=0.5\textwidth]{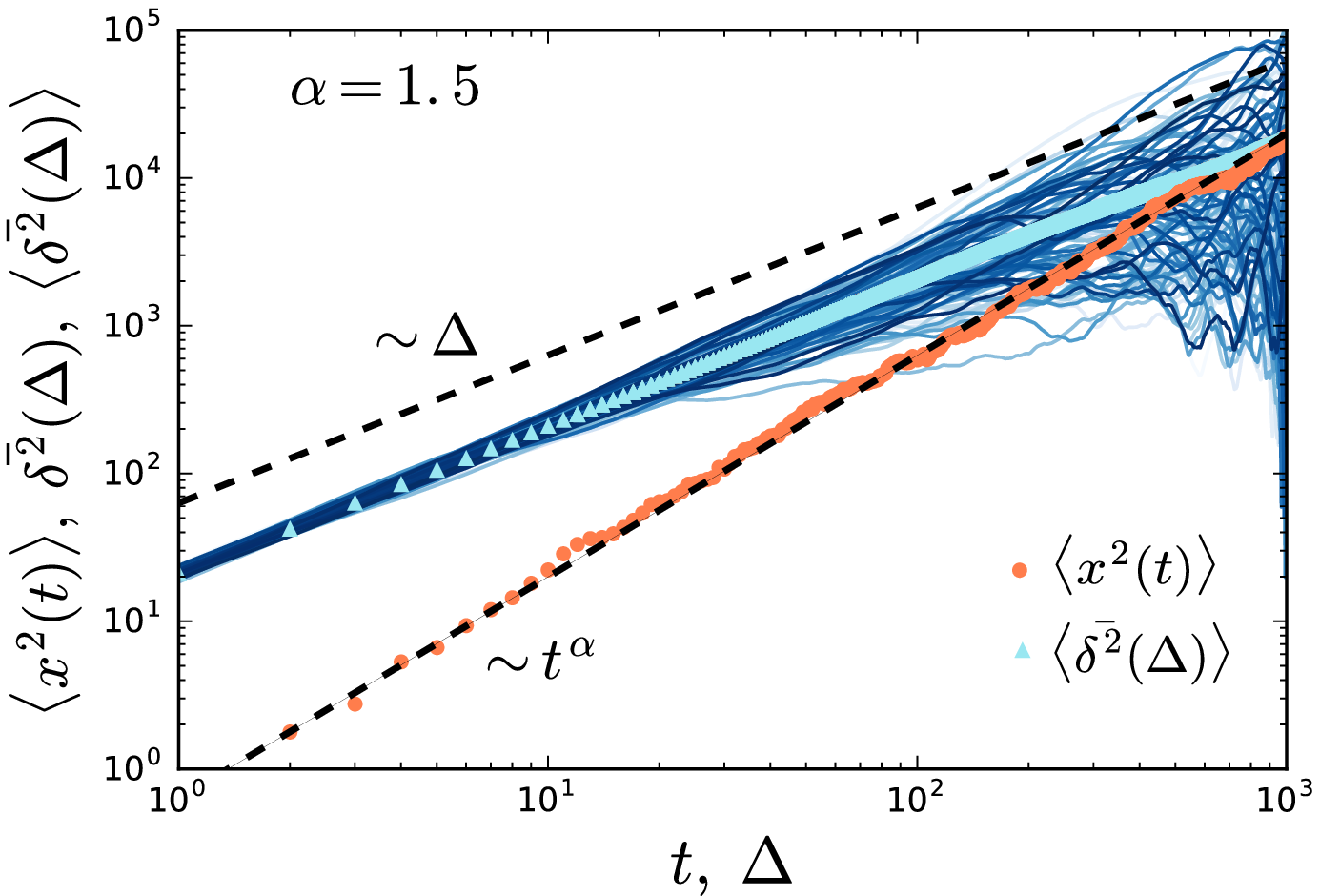}
\caption{SBM mean squared displacements (MSDs). For three different $\alpha$
values (subdiffusion with $\alpha=0.5$, top; normal diffusion with $\alpha=1$,
middle; superdiffusion with $\alpha=1.5$, bottom). Blue lines represent time
averaged MSDs $\overline{\delta^2(\Delta)}$ for individual trajectories. For
small $\Delta\ll T$ the individual $\overline{\delta^2(\Delta)}$ are fully
reproducible, while for longer lag time $\Delta$ the statistics becomes worse
and the trajectory-to-trajectory spread is appreciable. The light blue line
represents the trajectory-average $\langle\overline{\delta^2(\Delta)}\rangle$
of the time averaged MSD while the orange line depicts the ensemble averaged
MSD $\langle x^2(t)\rangle$.}
\label{img01}
\end{figure}

Before we proceed, it is expedient to recall other salient properties of SBM. In
particular, its autocorrelation function can be readily calculated to give
\begin{equation}
\langle X_\alpha(t_1) X_\alpha(t_2) \rangle=2 K_\alpha[\min\{t_1,t_2\}]^\alpha.
\label{SBM3}
\end{equation}
Hence, the covariance of $X_{\alpha}(t)$ has essentially the same form as the one
for standard Brownian motion, except that the time variable is "scaled". A basic
quantity to analyse the behaviour of individual trajectories is the time-averaged
MSD of the time series $X_{\alpha}(t)$ in the time interval $[0,T]$ \cite{pccp},
\begin{equation}
\overline{\delta^2(\Delta)}=\frac{1}{T-\Delta}\int^{T-\Delta}_0\Big(X_{\alpha}(t
+\Delta)-X_{\alpha}(t)\Big)^2dt,
\end{equation}
and its ensemble-averaged counterpart, which, taking into account expression
(\ref{SBM3}), can be explicitly calculated as \cite{SBM_Jeon}
\begin{equation}
\left<\overline{\delta^2(\Delta)}\right>=\frac{2K_{\alpha}}{\alpha+1}\left(\frac{
T^{\alpha+1}-\Delta^{\alpha+1}}{T-\Delta}-(T-\Delta)^{\alpha}\right).
\end{equation}
In the limit $\Delta\ll T$ we thus find $\langle\overline{\delta^2(\Delta)}\rangle
\sim2K_{\alpha}\Delta/T^{1-\alpha}$, a behaviour fundamentally different from the
ensemble-averaged MSD (\ref{msd}), a feature of so-called weak ergodicity breaking:
$\langle\overline{\delta^2(\Delta)}\rangle\neq\langle X_{\alpha}^2(\Delta)\rangle$
\cite{pccp}. We display the behaviour of individual time-averaged MSDs $\overline{
\delta^2(\Delta)}$ in figure \ref{img01}, along with their ensemble average $\langle
\overline{\delta^2(\Delta)}\rangle$ and the standard MSD $\langle X_{\alpha}^2(t)
\rangle$. The non-ergodic behaviour of SBM is clearly highlighted by the different
slopes of $\langle\overline{\delta^2(\Delta)}\rangle$ and $\langle X_{\alpha}^2(t)
\rangle$.

Equipped with all necessary knowledge on the properties of SBM $X_{\alpha}(t)$,
we now turn to the question of interest here, the analysis of its single-trajectory
PSD. As $S(f,T)$ is a random variable, the most general information about its
properties is contained in the moment-generating function
\begin{equation}
\Phi_\lambda=\langle\exp\left(-\lambda S(f,T)\right)\rangle,\quad\lambda\geq0.
\label{PSD2}
\end{equation}
Once $\Phi_{\lambda}$ is determined, the PDF $P(S(f,T)=S)$ of the random variable
$S(f,T)$ can be simply derived from equation (\ref{PSD2}) by an inverse Laplace
transform with respect to the parameter $\lambda$. As we proceed to show below,
both $\Phi_\lambda$ and $P(S(f,T)=S)$ are entirely defined by the first two moments,
due to the Gaussian nature of the process $X_{\alpha}(t)$. The mean value, which
represents the standard time-dependent PSD, is given by 
\begin{equation}
\label{mu}
\mu(f,T)=\left\langle S(f,T)\right\rangle,
\end{equation}
while the variance of the random variable $S(f,T)$ obeys
\begin{equation}
\label{sigma}
\sigma^2(f,T)=\left\langle S^2(f,T)\right\rangle-\mu^2(f,T).
\end{equation}
The calculation of the exact explicit forms of the properties defined in equations
(\ref{PSD2}) to (\ref{sigma}) represents the chief goal of our work.

\section{Spectral analysis of individual trajectories of scaled Brownian motion}
\label{sec3}

\begin{figure}
\centering
\includegraphics[width=0.6\textwidth]{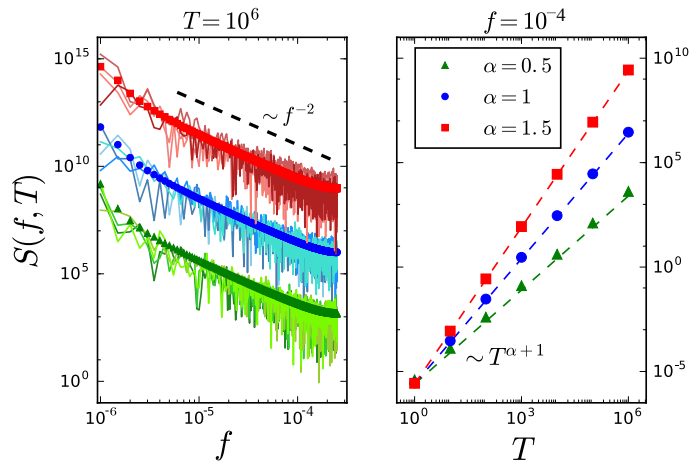}
\caption{Left: Single trajectory power spectra $S(f,T)$ for subdiffusion ($\alpha
=1/2$, top), normal diffusion ($\alpha=1$, middle), and superdiffusion ($\alpha=
3/2$, bottom) as function of frequency $f$. The thick lines represent the mean of
the simulated PSDs $S(f,T)$. The $1/f^2$ trend is indicated by the dashed line.
Right: mean of $S(f,T)$, averaged over individual trajectories, as function of
the observation time $T$.}
\label{powerspec}
\end{figure}

The single-trajectory PSDs $S(f,T)$ for four different sample trajectories for
the three anomalous diffusion exponents $\alpha=1/2$, $\alpha=1$, and $\alpha
=3/2$ are shown in figure \ref{powerspec}. While, naturally, we observe distinct
fluctuations within $S(f,T)$ and between different realisations, all data clearly
show a $S(f,T)\simeq1/f^2$-scaling. The right panel of figure \ref{powerspec}
demonstrates the apparent scaling of the trajectory-averaged single-trajectory
PSD as function of the observation time $T$ (ageing behaviour)---with the $T^{
1+\alpha}$-scaling derived below. We are now going to quantify these behaviours
in detail.

\subsection{Moment-generating function of the single-trajectory PSD}

We start from definition (\ref{PSD1}) of the single-trajectory PSD and rewrite
it in the form
\begin{eqnarray}
\label{new}
S(f,T)=\frac{1}{T}\int_0^T\int_0^T\cos(f(t_1-t_2))X_\alpha(t_1)X_\alpha(t_2)dt_1dt_2,
\end{eqnarray}
which is just a formal procedure since $X_\alpha(t)$ is a real-valued process. 
Relegating the intermediate steps of the derivation to \ref{app}, we eventually
find the exact result
\begin{eqnarray}
\Phi_\lambda&=&\left\langle\exp\left(-\frac{\lambda}{T}\int_0^T\int_0^T\cos\left(f
(t_1-t_2)\right)X_\alpha(t_1)X_\alpha(t_2)dt_1dt_2\right)\right \rangle\nonumber\\
&=&\frac{1}{\sqrt{1+2\mu\lambda+(2\mu^2-\sigma^2)\lambda^2}},
\label{momgen}
\end{eqnarray}
where $\mu$ and $\sigma^2$ are defined in equations (\ref{mu}) and (\ref{sigma}),
respectively. Result (\ref{momgen}) shows that the PDF of the single-trajectory PSD
for SBM is fully defined through its first and second moment, and that it has exactly
the same functional form as the results for Brownian motion and FBM derived in
\cite{PSD_BM,PSD_fBM}. As we have already remarked, this is a direct consequence of
the Gaussian nature of the parental process $X_{\alpha}(t)$ of SBM.

Inverting the Laplace transform with respect to $\lambda$ we obtain the PDF of the
random variable $S(f,T)$, 
\begin{equation}
P(S(f,T)=S)=\frac{1}{\mu\sqrt{2-\gamma^2}}\exp\left(-\frac{S}{\mu(2-\gamma^2)}\right)
I_0\left(\frac{\sqrt{\gamma^2-1}}{\mu(2-\gamma^2)}S\right),
\label{S_PDF}
\end{equation}
where $\gamma=\sigma/\mu$ is the coefficient of variation of the PDF of the
single-trajectory PSD for SBM, and $I_\nu$ is the modified Bessel function of the
first kind. The latter is known to be a distribution with heavier-than-Gaussian
tails.

\begin{figure}
\centering
\includegraphics[width=0.5\textwidth]{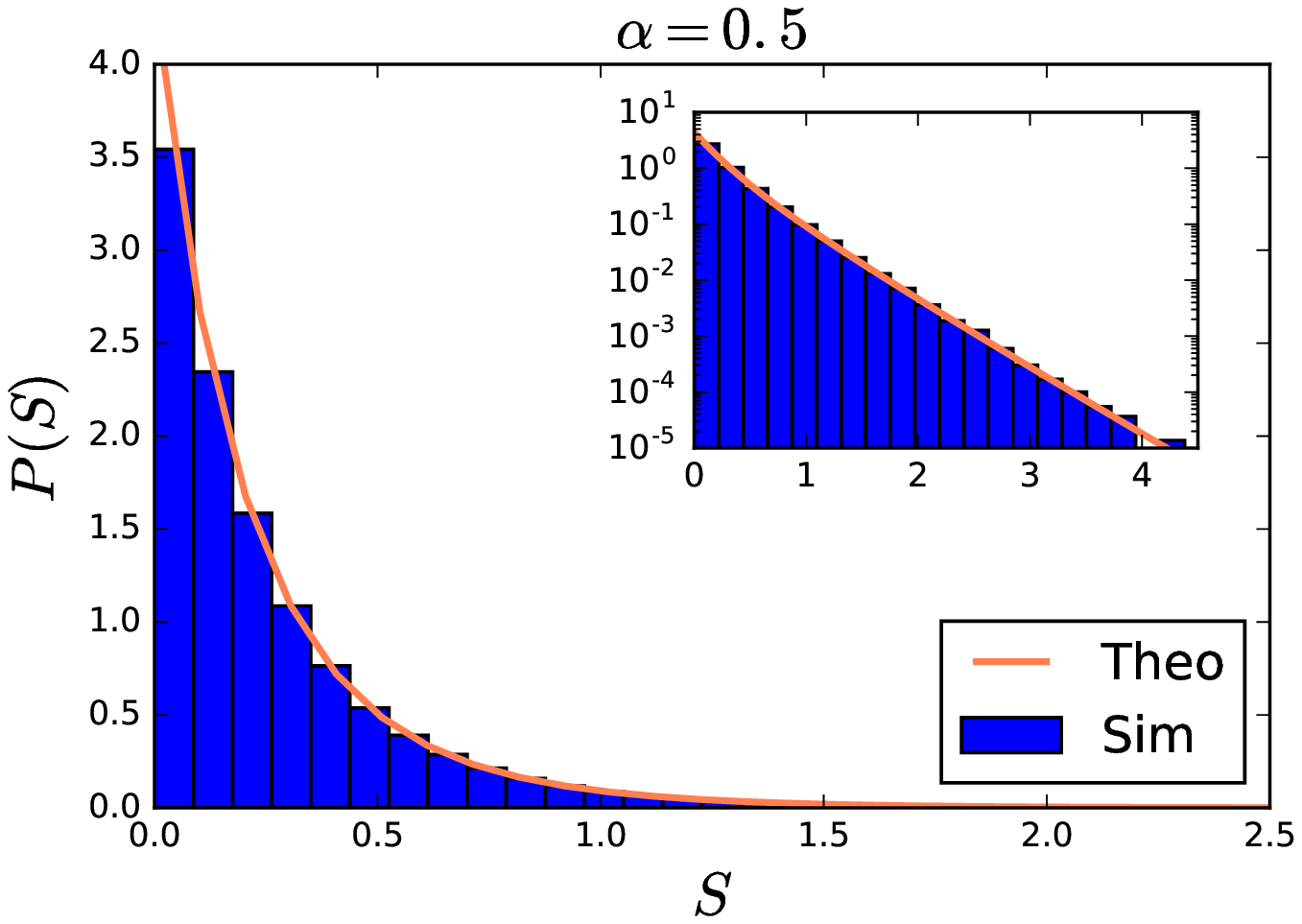}
\includegraphics[width=0.5\textwidth]{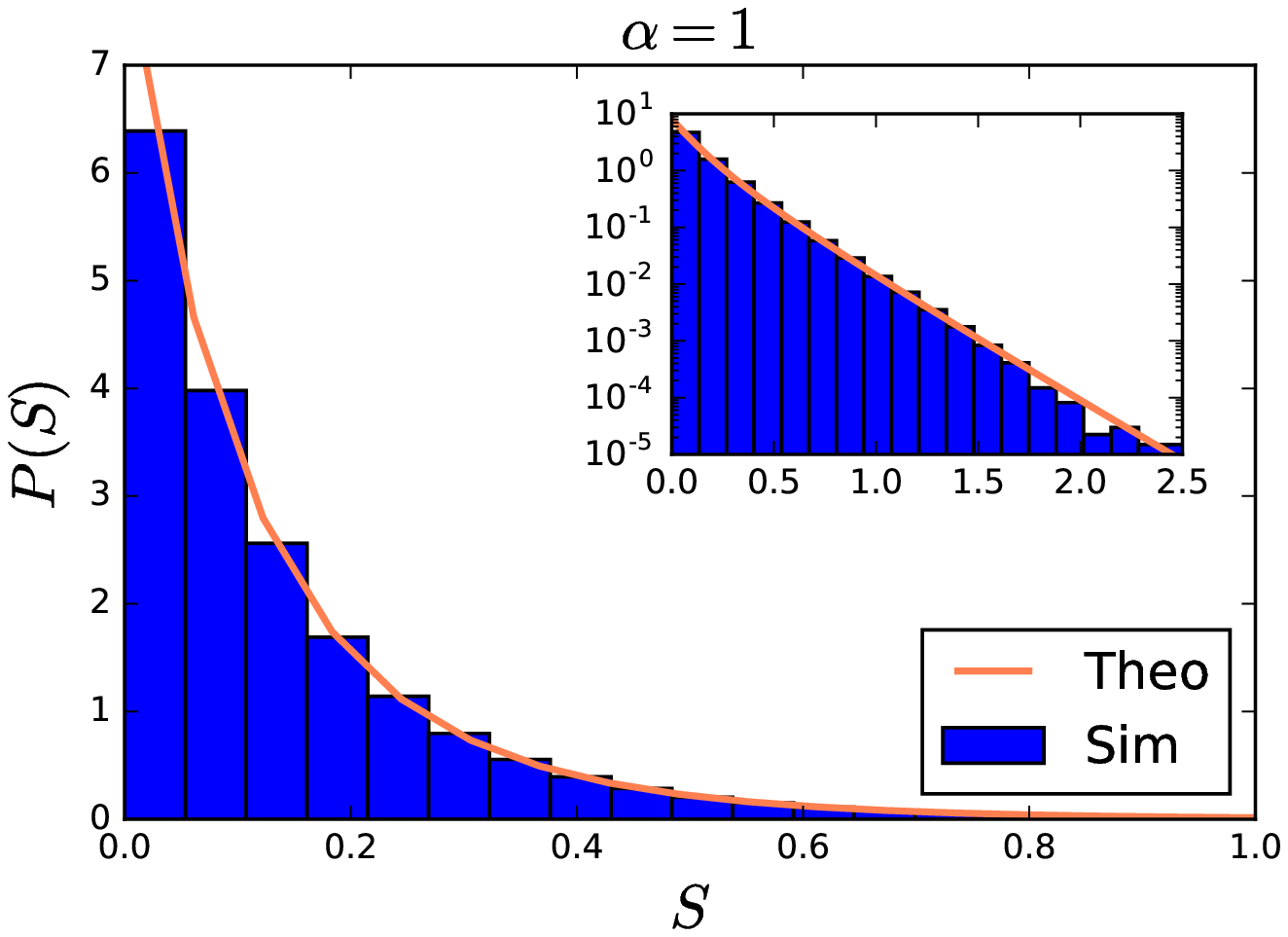}
\includegraphics[width=0.5\textwidth]{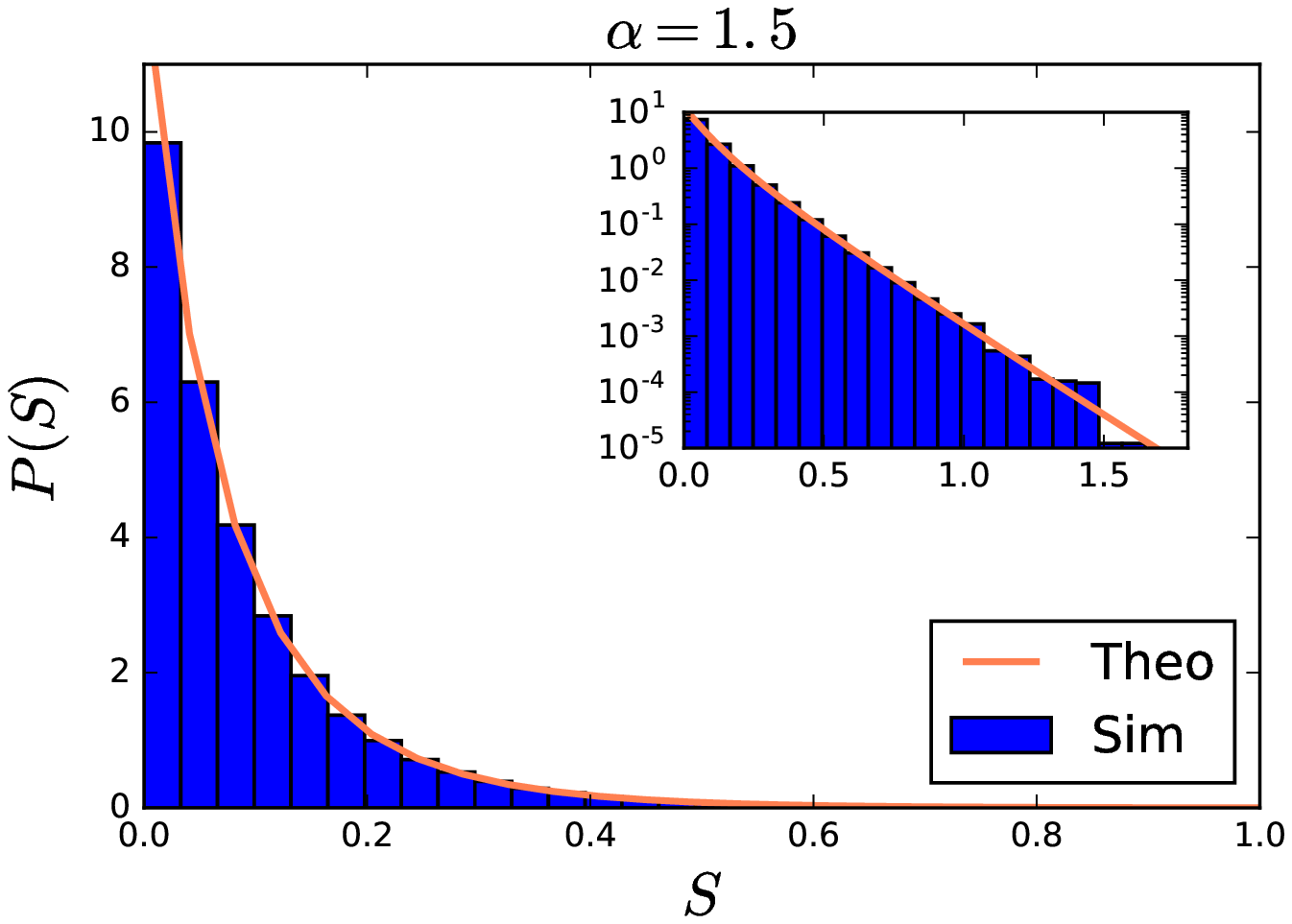}
\caption{Amplitude PDF $P(S(f,T)=S)$ of single-trajectory PSDs for different values
of the anomalous diffusion exponent: subdiffusive ($\alpha=1/2$, top), normal
diffusive ($\alpha=1$, middle), and superdiffusive ($\alpha=3/2$, bottom). In the
plots, "Theo" stands for the analytical result (\ref{S_PDF}), while "Sim" is the
histogram obtained from simulations, corresponding to averages over $10^6$
realisations for each $\alpha$. The insets report the same quantities on a
semi-logarithmic scale, demonstrating that the large-$S$ tail of the PDF in
equation (\ref{S_PDF}) is an exponential function. Analytical and numerical
results are in an excellent agreement.}
\label{img1}
\end{figure}

In figure \ref{img1} we present a comparison of the analytical result (\ref{S_PDF})
for $P(S(f,T)=S)$ with simulations. The agreement is excellent. The width of the
PDF $P(S(f,T)=S)$ becomes narrower for increasing $\alpha$ (note the different
scales on the axes). In particular, the insets show the exponential shape of the
PDF $P(S(f,T)=S)$ in the semi-logarithmic plots.

\subsection{Ensemble-averaged PSD}

We now proceed further and calculate the first moment of the PSD, defined in
equation (\ref{mu}). Recalling the expression for the autocorrelation function
(\ref{SBM3}) of SBM, we perform the integrations explicitly in \ref{app}, to
find the final expression
\begin{eqnarray}
\mu(f,T)=\frac{4K_\alpha T^{\alpha+1}}{fT}\left[\sin(fT)g_1\left(\frac{\alpha}{2},
fT\right)-\cos(fT)g_2\left(\frac{\alpha}{2},fT\right)
\right],
\label{mu6}
\end{eqnarray}
where we introduced the functions $g_1$ and $g_2$ defined in \ref{app}. It is
straightforward to check that for $\alpha =1$ equation (\ref{mu6}) yields the
standard expression of the PSD for Brownian motion,
\begin{equation}
\label{mu7}
\mu(f,T)\Big|_{\alpha=1}=\frac{4K_1}{f^2}\left(1-\frac{\sin(fT)}{fT}\right),
\end{equation}
where $K_1$ is the normal diffusion coefficient of dimensionality $\mathrm{cm}^2
/\mathrm{sec}$.

Next, we focus on the asymptotic behaviour of the general expression (\ref{mu})
in the limit $fT\to\infty$, which is equivalent to either the limit $f\to\infty$
with $T$ fixed, or vice versa. We get
\begin{equation}
\mu(f,T)\sim\frac{4K_\alpha T^{\alpha-1}}{f^2}\left\{1-\frac{\Gamma(\alpha +1)
\cos(fT-\frac{\pi\alpha}{2})}{(fT)^{\alpha}}\right\}.
\label{mu1}
\end{equation}
Interestingly, the $f$-dependence of the leading term is the same for \emph{any\/}
$\alpha$, in particular, it is equal to the one for Brownian motion. The fact that
we are not able to distinguish SBM from Brownian motion by just looking at the
frequency domain can lead, when analysing data, to the wrong conclusion that one
deals with standard Brownian motion. Only when we have sufficiently precise data
over a large frequency window, we could use the $\alpha$-dependent subleading term to
identify the anomalous diffusion exponent $\alpha$. The only explicit
$\alpha$-dependence in the leading order of expression (\ref{mu1}) is in the
ageing behaviour encoded by the dependence on $T^{\alpha-1}$ in the prefactor,
which therefore becomes a relevant behaviour to check.

The dependence on $\alpha$ of the ageing factor leads to the convergence of the
limit $T\to\infty$ in the subdiffusive case and to a divergence in the superdiffusive
case. A second interesting limit is given by the low-frequency limit $f=0$. In this
case we obtain 
\begin{equation}
\mu(f=0,T)=\frac{4K_\alpha T^{\alpha+1}}{(\alpha+1)(\alpha +2)}.
\label{mu8}
\end{equation}
This result represents the averaged squared area under the random curve
$X_{\alpha}(t)$.

\subsection{Variance and the coefficient of variation}

The variance of the single-trajectory PSD is defined in equation (\ref{sigma}).
It can be calculated exactly for arbitrary $\alpha$, $f$ and $T$, and the
details of the intermediate steps are presented in \ref{app}. Here we report
the asymptotic behaviour for $fT\to\infty$, reading
\begin{eqnarray}
\nonumber
\sigma^2(f,T)&\sim&\frac{16 K_\alpha^2T^{2\alpha-2}}{f^4}\left\{\frac{5}{4}
+\frac{\Gamma^2(\alpha+1)}{(fT)^{2\alpha}}+\frac{2\Gamma(\alpha+1)\cos^2(fT-
\frac{\pi\alpha}{2})}{2^{\alpha+2}(fT)^{\alpha}}\right.\\
\nonumber
&&-\frac{3\Gamma(\alpha+1)\cos(fT-\frac{\pi\alpha}{2})}{(fT)^{\alpha}}+
\frac{\Gamma^2(\alpha+1)\cos^2(fT-\frac{\pi\alpha}{2})}{(fT)^{2\alpha}}\\
&&-\left.\frac{\Gamma^2(\alpha+1)}{2^{\alpha+2}(fT)^{2\alpha}}[4\cos(fT)-1]
\right\}.
\label{var}
\end{eqnarray}
As for the mean value, the $f$-dependence of the leading term does not involve
$\alpha$, and it has the same scaling as Brownian motion. Similarly, the explicit
dependence on $\alpha$ of the frequence appears only in subleading order. Once
again, studying the leading frequency scaling only we are not able to distinguish
SBM from Brownian motion. Instead, we should pay attention to the ageing behaviour
of the amplitude.

We summarise the results for the mean and variance of the single-trajectory PSD
in the behaviour of the coefficient of variation, $\gamma$. It was shown that
for FBM this dimensionless factor plays the role of a delicate key criterion to
identifying anomalous diffusion. Namely $\gamma$ assumes three different values
in the limit of large frequency depending on whether we have sub-, normal or
superdiffusion, but independent of the precise value of $\alpha$. In the SBM
case, recalling the asymptotic results for the mean and variance in equations
(\ref{mu1}) and (\ref{var}) respectively, we obtain
\begin{equation}
\label{limit}
\gamma\sim\sqrt{5}/2, \quad f\to\infty
\end{equation}
for any $\alpha$. Moreover in the limit of $fT\to0$ we obtain
\begin{equation}
\sigma^2(f=0,T)=\frac{32K^2_\alpha T^{2\alpha+2}}{(\alpha +1)^2(\alpha+2)^2}
\quad\mbox{and}\quad\gamma\sim\sqrt{2}.
\label{var0}
\end{equation}
In this limit the moment-generating function simplifies and the probability
density function is the gamma distribution with scale $2\mu(f=0,T)$ and
shape parameter $1/2$.

\begin{figure}
\centering
\includegraphics[height=0.34\textwidth]{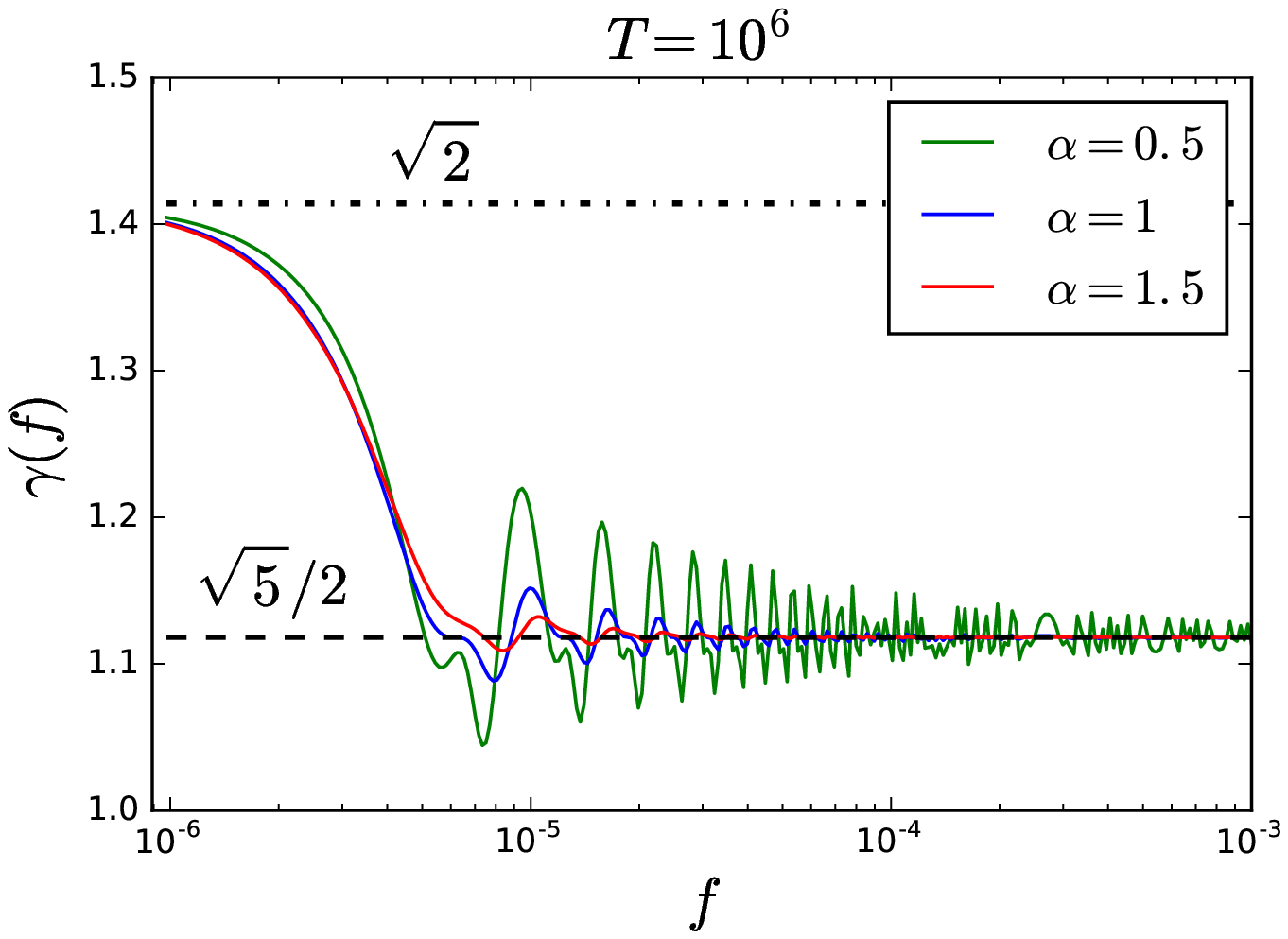}
\includegraphics[height=0.34\textwidth]{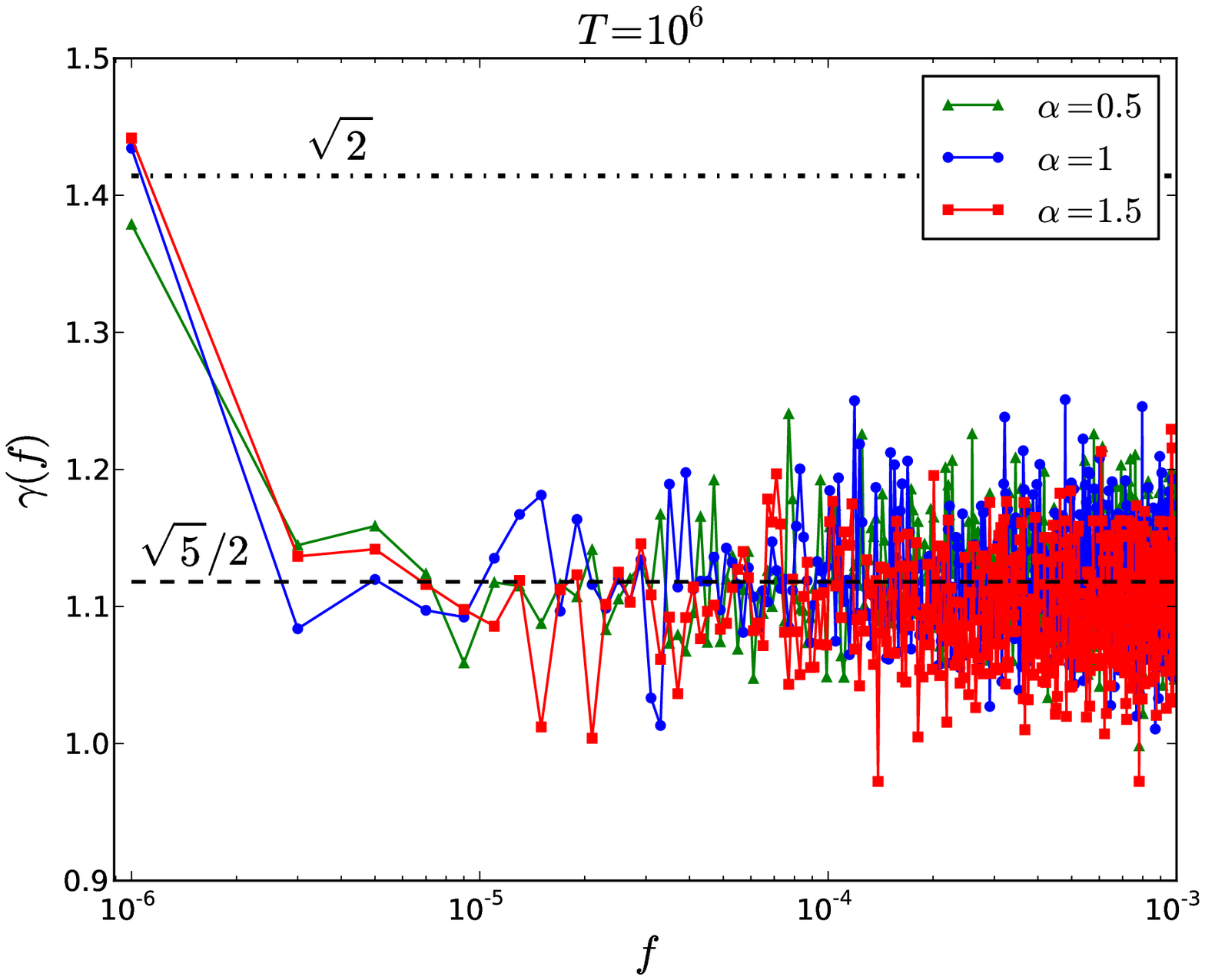}
\caption{Left panel: analytical behaviour of $\gamma$ for 3 different values
of $\alpha$ corresponding to sub-, normal and super-diffusion. Right panel:
$\gamma$ obtained from $10^3$ realisations of SBM, each consisting of $N=10^6$
time steps.}
\label{img2}
\end{figure}

In figure \ref{img2} analytical and numerical results for the coefficient of
variation $\gamma$ are shown. Analytically, in the case of subdiffusion we
observe heavier oscillation of $\gamma$ as function of the frequency, while
in the superdiffusive case the convergence to the limiting value (\ref{limit})
is faster. Such a distinction is not so clear in the numerics, where the
behaviour of $\gamma$ is essentially the same for the three different values
of $\alpha$, showing again the difficulties in differentiating SBM from Brownian
motion.

\subsection{Comparison with FBM}

The results obtained for SBM show both similarities and dissimilarities with
the ones for FBM reported in \cite{PSD_fBM}. In fact, both processes share
the same form for the PDF $P(x,t)$ in an infinite space and are therefore
often confused with one another in literature, see the caveats raised in
\cite{SBM_Jeon,pccp}. However, while both processes are obviously Gaussian,
FBM has stationary increments yet long-ranged, power-law noise correlations.
In contrast, SBM is non-stationary but driven by uncorrelated noise. After
our results above a natural question is whether in terms of the single-trajectory
PSD the two processes can be told apart.

For the frequency dependence of the single-trajectory PSD $S(f,T)$, and thus
also the mean $\mu(f,T)$, SBM shares the $1/f^2$ scaling with that of Brownian
motion for any value of the anomalous diffusion exponent $\alpha$ in the range
$0<\alpha<2$. Subdiffusive FBM, in contrast, exhibits a completely different
behaviour with the explicitly $\alpha$-dependent frequency scaling $1/f^{\alpha
+1}$. Moreover, while in the subdiffusive regime SBM shows the ageing dependence
$\mu\simeq T^{\alpha-1}$, FBM is independent of $T$. Thus, SBM and FBM can be
told apart quite easily from both $f$ and $T$ dependencies. In contrast, in
the superdiffusive regime the results for SBM and FBM are the same for the
functional behaviours with respect to both $f$ and $T$, and the processes
therefore cannot be told apart from each other by use of the single-trajectory
PSD or its mean. However, indeed there exists a difference when we consider the
coefficient of variation $\gamma$. Namely, for SBM $\gamma$ always converges to
the value $\gamma\sim\sqrt{5}/2$ at high frequencies, the value shared with
Brownian motion. FBM, in contrast, assumes three distinct values in the high
frequency limit: $\gamma\sim1$ for subdiffusion ($0<\alpha<1$), $\gamma\sim
\sqrt{5}/2$ for normal diffusion ($\alpha=1$), and $\gamma\sim\sqrt{2}$ for
superdiffusion ($1<\alpha<2$). These predictions are confirmed by numerical
and experimental data \cite{PSD_BM,PSD_fBM}. The coefficient of variation
therefore provides a suitable tool to distinguish SBM from FBM. We note that
it is not necessary that the value of $\gamma$ has fully converged within the
frequency window probed by experiment or simulation. It is sufficient to see
from the data whether a clear trend for a departure from the value $\sqrt{5}/2$
assumed by Brownian motion and SBM.

\section{Conclusions}
\label{conc}

We here studied the spectral content of SBM, a standard model for anomalous
diffusion which is Markovian but non-stationary, in terms of the single-trajectory
PSD and its full distribution. From analytical and numerical analyses we showed
that the frequency dependence has the invariant scaling form $\sim1/f^2$, fully
independent of the anomalous scaling exponent $\alpha$. We also showed that the
coefficient of variation for any $\alpha$ practically has the same frequency
dependence as for Brownian motion. The main difference between SBM and Brownian
motion is the ageing behaviour of single-trajectory PSD and its mean, that is,
their dependence on the observation time $T$.

FBM, in contrast, has stationary increments yet is non-Markovian due to its
power-law correlated driving noise. Both FBM and SBM are Gaussian in nature,
and we found both emerging similarities and dissimilarities. For both sub-
and superdiffusion the coefficient of variation for FBM provides different
values from SBM. In addition, subdiffusive FBM is non-ageing but has an
$\alpha$-dependent frequency scaling of the single-trajectory PSD. The
situation is different in the superdiffusive regime: here the frequency
dependence and the ageing behaviour of the single-trajectory PSD for FBM is
the same as for SBM, leaving the coefficient of variation as the only way to
distinguish the two processes from each other. Taking
together all observables, we conclude that the single-trajectory PSD is
able to distinguish SBM, FBM, and normal Brownian motion. We note that the
PDF of the single-trajectory PSD is, however, the same for all three cases.

The results reported here for SBM adds an important additional piece to the
development of a complete picture for single-trajectory PSD analysis of
modern single particle tracking data. We demonstrated that it is a suitable
tool to identify the anomalous scaling exponent $\alpha$ from an individual
particle trajectory $X_\alpha(t)$. Moreover, within the Gaussian processes studied
so far, the single-trajectory PSD framework allows one to tell the different
processes apart from each other, and is thus an outstanding physical observable,
providing complementary information to the (more) standard analyses in terms
of ensemble and time averaged MSDs.

\ack

RM acknowledges funding through grants ME 1535/6-1 and ME 1535/7-1 of Deutsche
Forschungsgemeinschaft (DFG), as well as through an Alexander von Humboldt
Polish Honorary Research Scholarship from the Polish Science Foundation.

\appendix

\section{Mean single-trajectory PSD}
\label{app}

Recalling definition (\ref{mu1}) we have
\begin{eqnarray}
\fl \mu (f,T) & = & \frac{1}{T} \int_0^T dt_1 \int_0^T dt_2 \cos (f (t_1-t_2)) 2 K_\alpha \min(t_1,t_2)^\alpha \nonumber \\
\fl & = & \frac{2K_\alpha}{T} \int_0^T dt_1\left\{ \int_0^{t_1} dt_2 \cos  (f (t_1-t_2)) t_2^\alpha + \int_{t_1}^T dt_2 \cos  (f (t_1-t_2)) t_1^\alpha \right\} \nonumber \\
\fl &=&  \frac{2K_\alpha}{T} \left\{ \int_0^T dt_1 \int_0^{t_1} dt_2  \left[ \cos(f t_1) \cos(f t_2) +\sin(ft_1) \sin (f t_2) \right] t_2^\alpha \right. \nonumber \\
\fl & & \left. + \int_0^T dt_1 \int_{t_1}^T dt_2 \left[ \cos(f t_1) \cos(f t_2) + \sin (f t_1) \sin (f t_2) \right] t_1^\alpha \right\} \nonumber \\
\fl &=&  \frac{2K_\alpha}{T}  \left\{ I_1+ I_2 + I_3 + I_4 \right\}.
\end{eqnarray}
We focus on the explicit calculation of each integral individually, starting with
\begin{eqnarray}
\nonumber
\fl I_1&=&\int_0^Tdt_1\int_0^{t_1}dt_2\cos(f t_1)\cos(ft_2)t_2^\alpha=\int_0^Tdt_1
\cos(ft_1)\int_0^{t_1}dt_2\cos(ft_2)t_2^\alpha\\
\nonumber
\fl & =&T^{\alpha+2}\int_0^1dy\cos(\omega y)y^{\alpha+1}\int_0^1dz\cos(\omega yz)
z^\alpha\\
\fl &=&T^{\alpha+2}\int_0^1dy\cos(\omega y)y^{\alpha+1}g_1(\frac{\alpha}{2},\omega y), 
\label{mu3}
\end{eqnarray} 
where $\omega=fT$ and 
\begin{eqnarray}
\nonumber
g_1(\alpha,\omega)&=&\int_0^1\tau^{2\alpha}\cos(\omega\tau)d\tau\\
\nonumber
&=&-\frac{\Gamma(2\alpha+1)\sin(\pi\alpha)}{\omega^{2\alpha+1}}\\
&&-\frac{i}{2\omega^{2\alpha+1}}\left(\e^{i\pi\alpha}\Gamma(2\alpha+1,-i\omega)-
\e^{-i\pi\alpha}\Gamma(2\alpha+1,i\omega)\right).  
\label{g1}
\end{eqnarray}
Similarly for the second integral we obtain 
\begin{eqnarray}
\nonumber
I_2&=&\int_0^Tdt_1\sin(ft_1)\int_0^{t_1}dt_2\sin(ft_2)t_2^\alpha\\
&=&T^{\alpha+2}\int_0^1dy\sin(\omega y)y^{\alpha+1}g_2(\frac{\alpha}{2},\omega y),
\label{mu4}
\end{eqnarray}
where
\begin{eqnarray}
\nonumber
g_2(\alpha,\omega)&=&\int_0^1\tau^{2\alpha}\sin(\omega\tau)d\tau\\
\nonumber
&=&\frac{\Gamma(2\alpha+1)\cos(\pi\alpha)}{\omega^{2\alpha+1}}\\
&&-\frac{1}{2\omega^{2\alpha+1}}\left(\e^{i\pi\alpha}\Gamma(2\alpha+1,-i\omega)+
\e^{-i\pi\alpha}\Gamma(2\alpha+1,i\omega)\right). 
\label{g2}
\end{eqnarray}
Plugging in the explicit expressions of $g_1(\alpha,\omega)$ and $g_2(\alpha,
\omega)$ and working out the integrals we arrive at
\begin{eqnarray}
\nonumber
\fl I_1&=&T^{\alpha+2}\left\{-\frac{\Gamma(\alpha+1)\sin(\pi\alpha/2)\sin(\omega)}{
\omega^{\alpha+2}}-\frac{\Gamma(\alpha+1)\cos(\pi\alpha/2)}{(2\omega)^{\alpha+2}}
\right.\\
\nonumber
\fl&&-i\frac{\sin(\omega)}{2\omega^{\alpha+2}}\left[\e^{i\pi\alpha/2}\Gamma(\alpha+1,
-i\omega-\e^{-i\pi\alpha/2} \Gamma(\alpha+1,i\omega)\right]\\
\fl&&\left.+\frac{1}{2(2\omega)^{\alpha+2}}\left[\e^{i\pi\alpha/2}\Gamma(\alpha+1,-2i
\omega)+\e^{-i\pi\alpha/2}\Gamma(\alpha+1,2i\omega)\right]\right\},\\
\nonumber
\fl I_2&=&T^{\alpha+2}\left\{-\frac{\Gamma(\alpha+1)\cos(\pi\alpha/2)\cos(\omega)}{
\omega^{\alpha+2}}+\frac{\Gamma(\alpha+1)\cos(\pi\alpha/2)}{(2\omega)^{\alpha+2}}
\right.\\
\nonumber
\fl&&+\frac{\cos(\omega)}{2\omega^{\alpha+2}}\left[\e^{i\pi\alpha/2}\Gamma(\alpha+1,-i
\omega)+\e^{-i\pi\alpha/2}\Gamma(\alpha+1,i\omega)\right]\\
\fl&&\left.-\frac{1}{2(2\omega)^{\alpha+2}}\left[\e^{i\pi\alpha/2}\Gamma(\alpha+1,-2i
\omega)+\e^{-i\pi\alpha/2}\Gamma(\alpha+1,2i\omega)\right]\right\}.
\label{mu5}
\end{eqnarray}
The last two integrals are given by
\begin{eqnarray}
\nonumber
\fl I_3&=&\int_0^Tdt_1\cos(ft_1)t_1^\alpha\int_{t_1}^Tdt_2\cos(ft_2)\\
\fl&=&\frac{T^{\alpha+2}}{\omega}\left\{\sin(\omega)g_1(\frac{\alpha}{2},\omega)-
\frac{1}{2}\int_0^1dyy^\alpha\sin(2\omega y)\right\},\\
\nonumber
\fl I_4&=&\int_0^Tdt_1\sin(ft_1)t_1^\alpha\int_{t_1}^Tdt_2\sin(ft_2)\\
\fl&=&\frac{T^{\alpha+2}}{\omega}\left\{-\cos(\omega)g_2(\frac{\alpha}{2},\omega)
+\frac{1}{2}\int_0^1dyy^\alpha\sin(2\omega y)\right\},
\end{eqnarray}
so that we finally obtain
\begin{eqnarray}
\nonumber
\mu(f,T)&=&2K_\alpha T^{\alpha +1}\left\{-\frac{\Gamma(\alpha +1)}{\omega^{
\alpha+2}}\cos\left(\omega-\frac{\pi\alpha}{2}\right)\right.\\
\nonumber
&&+\frac{\cos(\omega-\frac{\pi\alpha}{2})}{2\omega^{\alpha+2}}\left[\Gamma(
\alpha+1,i\omega)+\Gamma(\alpha+1,-i\omega)\right]\\
\nonumber
&&+\frac{i\sin(\omega-\frac{\pi\alpha}{2})}{2\omega^{\alpha+2}}\left[\Gamma
(\alpha+1,i\omega)-\Gamma(\alpha+1,-i\omega)\right]\\
&&\left.+\frac{1}{\omega}\left[\sin(\omega)g_1(\frac{\alpha}{2},\omega)-\cos
(\omega)g_2(\frac{\alpha}{2},\omega)\right]\right\},
\end{eqnarray}
which can be simplified to the form (\ref{mu6}).

\section{Moment-generating function of the single-trajectory PSD}

The moment-generating function is calculated as
\begin{eqnarray}
\fl\Phi_\lambda&=&\left\langle\exp\left\{-\lambda S(f,T)\right\}\right\rangle=\left
\langle\exp\left\{-\frac{\lambda}{T}\int_0^T\int_0^Tdt_1dt_2\cos(f(t_1-t_2))X_
\alpha(t_1)X_\alpha(t_2)\right\}\right\rangle\nonumber\\
\fl&=&\left\langle\exp\left\{-\frac{\lambda}{T}\left[\int_0^Tdt\cos(ft)X_\alpha(t)
\right]^2-\frac{\lambda}{T}\left[\int_0^Tdt\sin(ft)X_\alpha(t)\right]^2\right\}
\right\rangle\nonumber\\
\fl&=&\frac{T}{4\pi\lambda}\int_{-\infty}^{+\infty}dz_1\int_{-\infty}^{+\infty}dz_2
\exp\left(-T\frac{z_1^2+z_2^2}{4\lambda}\right)\nonumber\\
\fl&&\times\left\langle\exp\left\{iz_1\int_0^Tdt\cos(ft)X_\alpha(t)+iz_2\int_0^Tdt
\sin(ft)X_\alpha(t)\right\}\right\rangle\nonumber\\
\nonumber
\fl&=&\frac{T}{4\pi\lambda}\int_{-\infty}^{+\infty}dz_1\int_{-\infty}^{+\infty}dz_2
\exp\left(-T\frac{z_1^2+z_2^2}{4\lambda}\right)\\
\fl&&\times\left\langle\exp\left\{i\int_0^Tdt\xi(t)(z_1Q_1+z_2Q_2)\right\}\right
\rangle,
\label{mom_gen1}
\end{eqnarray}
where we used the identity $\exp(-b^2/4a)=\sqrt{a/\pi}\int_{-\infty}^{+\infty}dx
\exp(-ax^2+ibx)$ and we defined
\begin{eqnarray}
Q_1&=&\sqrt{2D_\alpha(t)}\left(\frac{\sin(fT)}{f}-\frac{\sin(ft)}{f}\right),
\nonumber \\
Q_2&=&\sqrt{2D_\alpha(t)}\left(\frac{\cos(ft)}{f}-\frac{\cos(fT)}{f}\right).
\end{eqnarray}
We can now average over the exponential of Gaussian variable and obtain
\begin{eqnarray}
\Phi_\lambda&=&\frac{T}{4\pi\lambda}\int_{-\infty}^{+\infty}dz_1\int_{-\infty}^{
\infty}dz_2\exp\left(-T\frac{z_1^2+z_2^2}{4\lambda}\right)\nonumber\\
&&\times\exp\left(-\frac{1}{2}\int_0^Tdtz_1^2Q_1^2-\frac{1}{2}\int_0^Tdtz_2^2Q_2^2
-\int_0^Tdtz_1z_2Q_1Q_2\right)\nonumber\\
&=&\left[1+\frac{4\lambda}{T}\left(\frac{A+B}{2}\right)+\left(\frac{4\lambda}{T}
\right)^2\frac{AB-C^2}{4}\right]^{-1/2},
\end{eqnarray} 
where for the last equality we used the identity $\int_{-\infty}^{+\infty}dz_1\int_{
-\infty}^{+\infty}dz_2\exp(-\alpha z_1^2+\beta z_2^2-2\gamma z_1z_2)=\pi(\alpha
\beta-\gamma^2)^{-1/2}$ and we defined
\begin{eqnarray}
\fl A&=&\int_0^TdtQ_1^2=\frac{2K_\alpha T^{\alpha+2}}{\omega}\left[2\sin\omega
g_1(\frac{\alpha}{2},\omega)-g_2(\frac{\alpha}{2},2\omega)\right],\nonumber\\
\fl B&=&\int_0^TdtQ_2^2=\frac{2K_\alpha T^{\alpha+2}}{\omega}\left[g_2(\frac{
\alpha}{2},2\omega)-2\cos\omega g_2(\frac{\alpha}{2},\omega)\right],\nonumber\\
\fl C&=&\int_0^TdtQ_1Q_2=\frac{2K_\alpha T^{\alpha+2}}{\omega}\left[\sin\omega
g_2(\frac{\alpha}{2},\omega)-\cos\omega g_1(\frac{\alpha}{2},\omega)+g_1(\frac{
\alpha}{2},2\omega)\right].
\end{eqnarray}
It is possible to show that 
\begin{eqnarray}
\frac{(A+B)}{T}&=&\mu(f,T),\label{rel1}\\
\frac{4(AB-C^2)}{T^2}&=&2\mu^2(f,T)-\sigma^2(f,T).\label{rel2}
\end{eqnarray}
Relations (\ref{rel1}) and (\ref{rel2}) allows us to rewrite the moment-generating
function as
\begin{equation}
\Phi_\lambda=\left[1+2\mu\lambda+(2\mu^2-\sigma^2)\lambda^2\right]^{-1/2}.
\end{equation}

\section{Variance of the single-trajectory PSD}

In order to obtain the PSD variance, given in (\ref{sigma}) we first focus on the
calculation of the second moment,
\begin{eqnarray}
\nonumber
\fl\langle S^2(f,T)\rangle&=&\frac{1}{T^2}\left\langle\int_0^T\int_0^Tdt_1dt_2\cos
(f(t_1-t_2))X_\alpha(t_1)X_\alpha(t_2)\right.\\
&&\times\left.\int_0^T\int_0^Tdt_3dt_4\cos(f(t_3-t_4))X_\alpha(t_3)X_\alpha(t_4)
\right\rangle\nonumber\\
\nonumber
\fl&=&\frac{1}{T^2}\int_0^T\int_0^T\int_0^T\int_0^Tdt_1dt_2dt_3dt_4\cos(f(t_1-t_2))
\cos(f (t_3-t_4))\\
&&\times\langle X_\alpha(t_1)X_\alpha(t_2)X_\alpha(t_3)X_\alpha(t_4)\rangle.
\label{var2}
\end{eqnarray}
Following the Wick/Isserlis theorem we have
\begin{eqnarray}
\nonumber
\langle X_\alpha(t_1)X_\alpha(t_2)X_\alpha(t_3)X_\alpha(t_4)\rangle&=&\langle
X_\alpha(t_1)X_\alpha(t_2)\rangle\langle X_\alpha(t_3)X_\alpha(t_4)\rangle\\
\nonumber
&&+\langle X_\alpha(t_1)X_\alpha(t_3)\rangle\langle X_\alpha(t_2)X_\alpha(t_4)
\rangle\\
&&+\langle X_\alpha(t_1)X_\alpha(t_4)\rangle\langle X_\alpha(t_3)X_\alpha(t_2)
\rangle.
\label{var3}
\end{eqnarray}
This allows us to rewrite (\ref{var2}) as
\begin{eqnarray}
\nonumber
\fl\langle S^2(f,T)\rangle&=&\frac{1}{T^2}\left\{\left[\int_0^T\int_0^Tdt_1dt_2
\cos(f(t_1-t_2))\langle X_\alpha(t_1)X_\alpha(t_2)\rangle\right]^2\right.\\
\nonumber
\fl&&+\int_0^T\int_0^T\int_0^T\int_0^Tdt_1dt_2dt_3dt_4\cos(f(t_1-t_2))\cos(f(
t_3-t_4))\\
\fl&&\times\langle X_\alpha(t_1)X_\alpha(t_3)\rangle\langle X_\alpha(t_2)X_
\alpha(t_4)\rangle\nonumber\\
\nonumber
\fl&+&\int_0^T\int_0^T\int_0^T\int_0^Tdt_1dt_2dt_3dt_4\cos(f(t_1-t_2))\cos
(f(t_3-t_4))\\
\fl&&\times\left.\langle X_\alpha(t_1)X_\alpha(t_4)\rangle\langle X_\alpha(t_3)X_
\alpha(t_2)\rangle\right\}\nonumber\\
\fl&=&\mu^2(f,T)+\frac{4K_\alpha^2}{T^2}\left\{\int_0^T\int_0^T\int_0^T\int_0^T
dt_1dt_2dt_3dt_4\cos(f(t_1-t_2))\cos(f (t_3-t_4))\right.\nonumber\\
\nonumber
\fl&&\times\min(t_1,t_3)^\alpha\min(t_2,t_4)^\alpha\\
&&+\int_0^T\int_0^T\int_0^T\int_0^Tdt_1dt_2dt_3dt_4\cos(f(t_1-t_2))\cos(f(t_3-t_4))
\nonumber\\
\fl&&\times\min(t_1,t_4)^\alpha\min(t_2,t_3)^\alpha\Big\}.
\label{var4}
\end{eqnarray}
The variance is thus given by
\begin{eqnarray}
\nonumber
\fl\sigma^2(f,T)&=&\frac{8K_\alpha^2}{T^2}\int_0^T\int_0^T\int_0^T\int_0^Tdt_1dt_2
dt_3dt_4\cos(f(t_1-t_2))\cos(f(t_3-t_4))\\
\fl&&\times\min(t_1,t_3)^\alpha\min(t_2,t_4)^\alpha\nonumber\\
\fl&=&\frac{8K_\alpha^2}{T^2}\left\{\int_0^T\int_0^Tdt_1dt_2\cos(f(t_1-t_2))\int_0
^{t_1}dt_3\int_0^{t_2}dt_4\cos(f(t_3-t_4))t_3^\alpha t_4^\alpha \right.\nonumber\\
\fl&&+2\int_0^T\int_0^Tdt_1dt_2\cos(f(t_1-t_2))\int_{t_1}^Tdt_3\int_0^{t_2}dt_4\cos
(f(t_3-t_4))t_1^\alpha t_4^\alpha\nonumber\\
\fl&&+\left.\int_0^T\int_0^Tdt_1dt_2\cos(f(t_1-t_2))\int_{t_1}^Tdt_3\int_{t_2}^T
dt_4\cos(f(t_3-t_4))t_1^\alpha t_2^\alpha\right\}\nonumber\\
\fl&=&\frac{8K_\alpha^2}{T^2}\left\{I_5+2I_6+I_7\right\}.
\label{var5}
\end{eqnarray}
Following the same procedure used above for calculating the mean we can show that
the integrals are given by
\begin{eqnarray}
\fl I_5&=&I_1^2+I_2^2+I_8^2+I_9^2,\\
\nonumber
\fl I_6&=&\frac{T^{\alpha +2}}{\omega}\left\{I_1\left[\sin(\omega)g_1\left(\frac{
\alpha}{2},\omega\right)-\frac{g_2\left(\frac{\alpha}{2},2\omega\right)}{2}\right]
\right.\\
\fl&&+I_8\left[\frac{1}{2(\alpha+1)}+\frac{g_1(\frac{\alpha}{2},2\omega)}{2}-\cos
(\omega)g_1\left(\frac{\alpha}{2},\omega\right)\right]\nonumber\\
\nonumber
\fl&&+I_9\left[-\frac{1}{2(\alpha+1)}+\frac{g_1(\frac{\alpha}{2},2 \omega)}{2}+
\sin(\omega)g_2\left(\frac{\alpha}{2},\omega\right)\right]\\
\fl&&\left.+I_2\left[-\cos(\omega)g_2\left(\frac{\alpha}{2},\omega\right)+\frac{
g_2(\frac{\alpha}{2},2\omega)}{2}\right]\right\},\\
\fl I_7&=&\frac{T^{\alpha +2}}{\omega^2}\left\{\frac{1}{2(\alpha+1)^2}+g_1\left(
\frac{\alpha}{2},\omega\right)^2+g_2\left(\frac{\alpha}{2},\omega\right)^2+\frac{
g_1(\frac{\alpha}{2},2 \omega)^2}{2}+\frac{g_2(\frac{\alpha}{2},2\omega)^2}{2}
\right.\nonumber\\
\fl&&-\sin(\omega)\left[\frac{g_2(\frac{\alpha}{2},\omega)}{\alpha+1}+g_1\left(
\frac{\alpha}{2},\omega\right)g_2\left(\frac{\alpha}{2},2\omega\right)-g_2\left(
\frac{\alpha}{2},\omega\right)g_1\left(\frac{\alpha}{2},2\omega\right)\right]
\nonumber\\
\fl&&-\left.\cos(\omega)\left[\frac{g_1(\frac{\alpha}{2},\omega)}{\alpha+1}+g_1
\left(\frac{\alpha}{2},\omega\right)g_1\left(\frac{\alpha}{2},2\omega\right)+
g_2\left(\frac{\alpha}{2},\omega\right)g_2\left(\frac{\alpha}{2},2\omega\right)
\right]\right\},
\label{var6}
\end{eqnarray}
where $I_1$ and $I_2$ are defined in (\ref{mu5}) and 
\begin{eqnarray}
\nonumber
\fl I_8&=& T^{\alpha+2}\left\{\frac{\Gamma(\alpha+1)\cos(\pi\alpha/2)\sin(\omega)}{
\omega^{\alpha+2}}-\frac{\Gamma(\alpha+1)\sin(\pi\alpha/2)}{(2\omega)^{\alpha+2}}
\right.\\
\nonumber
\fl&&-\frac{\sin(\omega)}{2\omega^{\alpha+2}}\left[\e^{i\pi\alpha/2}\Gamma(\alpha+1,
-i\omega)+\e^{-i\pi\alpha/2}\Gamma(\alpha+1,i\omega)\right]\\
\fl&&\left.-\frac{1}{2 \omega (\alpha+1)}- \frac{i}{2 (2\omega)^{\alpha+2}} \left[\e^{i\pi\alpha/2} \Gamma (\alpha +1, -2 i \omega) - \e^{-i\pi\alpha/2}\Gamma (\alpha +1, 2i \omega) \right] \right\},\\
\fl I_9&=&T^{\alpha+2}\left\{-\frac{\Gamma(\alpha+1)\sin(\pi\alpha/2)\cos(\omega)}{
\omega^{\alpha+2}}-\frac{\Gamma(\alpha+1)\sin(\pi\alpha/2)}{(2\omega)^{\alpha+2}}
\right.\\
\nonumber
\fl&&+i\frac{\cos(\omega)}{2\omega^{\alpha+2}}\left[\e^{i\pi\alpha/2}\Gamma(\alpha
+1,-i\omega-\e^{-i\pi\alpha/2}\Gamma(\alpha+1,i\omega)\right]\\
\fl&&+\left.\frac{1}{2 \omega (\alpha+1)}-\frac{i}{2(2\omega)^{\alpha+2}}\left[
\e^{i\pi\alpha/2}\Gamma(\alpha+1,-2i\omega)-\e^{-i\pi\alpha/2}\Gamma(\alpha+1,2i
\omega)\right]\right\}.
\end{eqnarray}

\section*{References}

\bibliographystyle{iopart-num}

\end{document}